\documentclass[preprint2]{aastex}

\usepackage{natbib}

\def\bdm{\begin{displaymath} }

\def\edm{\end{displaymath} }

\def\fr#1#2{{\textstyle{#1\over#2}}}

\begin{document}


\title{ Vibrationally Excited C$_4$H}

\author{Andrew~L.~Cooksy,\altaffilmark{1} C.~A.~Gottlieb,\altaffilmark{2} T.~C.~Killian,\altaffilmark{3} P.~Thaddeus,\altaffilmark{2,4}
Nimesh~A.~Patel,\altaffilmark{2} Ken~H.~Young,\altaffilmark{2} and M.~C.~McCarthy\altaffilmark{2}}

\altaffiltext{1}{Department of Chemistry and Biochemistry, San Diego State University, 5500 Campanile Drive, San Diego California 92182-1030}
\altaffiltext{2}{Harvard-Smithsonian Center for Astrophysics, 60 Garden Street, Cambridge, MA 02138}
\altaffiltext{3}{Department of Physics and Astronomy and the Rice Quantum Institute, Rice University, Houston, Texas 
77005-1892}
\altaffiltext{4}{School of Engineering and Applied Sciences, Harvard University, Cambridge, Massachusetts 02138}



\begin{abstract}  

Rotational spectra in four new excited vibrational levels of the linear carbon chain radical C$_4$H radical were observed in 
the millimeter band between 69 and 364~GHz in a low pressure glow discharge, and two of these were observed in a 
supersonic molecular beam between 19 and 38~GHz.
All have rotational constants within 0.4\% of the $X^2\Sigma^+$ ground vibrational state of C$_4$H and
were assigned to new bending vibrational levels, two each with $^2\Sigma$ and $^2\Pi$ vibrational symmetry.
The new levels are tentatively assigned to the $1\nu_6$ and  $1\nu_5$ bending vibrational modes (both with $^2\Pi$ symmetry), 
and the $1\nu_6 + 1\nu_7$ and $1\nu_5 + 1\nu_6$ combination levels ($^2\Sigma$ symmetry) on the basis of the derived spectroscopic constants, 
relative intensities in our discharge source, and published laser spectroscopic and quantum calculations.
Prior spectroscopic constants in the $1\nu_7$ and $2\nu_7$ levels were refined.
Also presented are interferometric maps of the ground state and the $1\nu_7$ level obtained with the SMA near 257~GHz which show that C$_4$H is present
near the central star in IRC+10216.
We found no evidence with the SMA for the new vibrationally excited levels of C$_4$H at a peak flux density averaged over a $3^{\prime\prime}$ 
synthesized beam of $\ge 0.15$~Jy/beam in the $294-296$ and $304-306$~GHz range, but it is anticipated that rotational lines in the new  levels might be observed 
in IRC+10216 when ALMA attains its full design capability.

\end{abstract}


\noindent     \hfill

\vspace{-0.5cm}

\section{Introduction}

Carbon-rich asymptotic giant branch (AGB) stars are a principal source of carbonaceous gas and dust in the interstellar medium, 
and some of the material from which planets are ultimately formed \citep{Molster2003,Ott2010}.
The dust is believed to form in the poorly explored inner envelope at $r \le 5R_{\star}$ where the temperatures and densities are high 
\citep[$T = 800 - 2000$~K and $n_{\rm{H_2}} = 10^9 - 10^{12}$~cm$^{-3}$;][]{Gail1988,Cherchneff2000,Agundez2006}.
During the past 40~years, astronomers have studied the chemistry and dynamics in the outer envelope of the prototypical carbon-rich 
AGB star \hfill\break
\noindent     \hfill
\vspace{0.025cm}

\noindent IRC+10216, but it is only in the last few years that  they have begun to observe a few small stable molecules in the inner envelope 
at high angular resolution with millimeter-wave interferometers.\footnote{Four small gaseous molecules have been 
identified in the dust formation zone from high resolution rotationally resolved spectra: CO \citep{Patel2009b}, HCN and HCCH 
\citep{Fonfria2008}, and HNC \citep{Cerni2013}.  Lacking a permanent electric dipole moment, HCCH was only observed in the IR,  
and silicon carbide dust (SiC) is inferred from a broad featureless band in the IR at $11\mu$m \citep{Speck1997}.} 


In the first wide band interferometric spectral line survey of IRC+10216, a large population of  narrow molecular emission lines 
were observed  whose expansion velocities ($V_{\rm{exp}} \sim 4$~km~s$^{-1}$) are about three times lower than the terminal velocity
\citep[14.5~km~s$^{-1}$;][]{Patel2009a}.  
Subsequent measurements at still higher angular resolution than the initial $3^{\prime\prime}$ confirmed that the emission from  
the narrow lines arises in the dust formation zone close to the star \citep{Patel2009b}.
In all, more than 200 narrow lines were observed in a 61~GHz wide spectral line survey in the 345~GHz band with the Submillimeter Array (SMA), of which the carriers of  approximately 50\%  remain unassigned \citep{Patel2011}.
When the same source was  observed at somewhat higher angular resolution and sensitivity with ALMA, the number of unassigned lines is over 10~times greater \citep[e.g., see the 1.9~GHz wide band near 267~GHz observed with a $0.6^{\prime\prime}$ diameter beam 
in ALMA Cycle~0 by][]{Cerni2013}.
Detection of such a large number of unassigned lines confirms that the supporting laboratory spectroscopy is far from complete, 
and the chemistry in the inner envelope of carbon-rich AGB stars  is not well understood.


After H$_2$ and CO, acetylene (HCCH) is the most abundant  species in the inner envelope of  IRC+10216   
\citep{Cherchneff2000,Fonfria2008}.
Both  on earth and in space, long carbon chain molecules are formed by
reactions of highly reactive  linear carbon chain  C$_n$H radicals with closed shell acetylenic molecules of the type  \hfill\break
\centerline  {C$_2$H + C$_n$H$_2$ $\rightarrow$ C$_{n+2}$H$_2$ + H}
\centerline   {C$_2$H + HC$_{2n+1}$N $\rightarrow$ HC$_{2n+3}$N +H} 
\noindent  \citep{Millar2000,Woods2003}.
\noindent 
Rotational lines of seven C$_n$H radicals (C$_2$H $\cdots$ C$_8$H) in the ground vibrational state have been studied extensively 
in the outer envelope of IRC+10216, and a few rotational lines have been observed in low-lying (bending) vibrational levels of C$_2$H, 
C$_3$H, C$_4$H, and C$_6$H at an angular resolution of $10^{\prime\prime} - 30^{\prime\prime}$ with single 
antennas (see Table~\ref{CnHspace}, and references therein). 
Chemical equilibrium calculations show that at temperatures near 1500~K, the abundance of the prototypical C$_2$H radical is comparable 
to that of HCCH  \citep{Yasuda2012}, but C$_n$H radicals have not yet been identified in the inner envelope.


In the course of measuring the rotational spectra of reactive acetylenic radicals, carbenes, and anions in a laboratory discharge through
HCCH and a rare gas, we have observed rotational lines in four vibrationally excited levels of C$_4$H that had not previously been studied.
Following the initial laboratory measurements of the C$_4$H radical in the ground vibrational state \citep{Gottlieb1983}, rotational lines were
observed in IRC+10216 and a laboratory discharge which were tentatively assigned to the $2\nu_7~(^2\Sigma)$ vibrational
level \citep{Guelin1987}.
At about the same time, \citet{Yamamoto} in a comprehensive laboratory study  measured the rotational spectra in the  $\nu_7~(^2\Pi)$ and $2\nu_7~(^2\Delta)$ excited vibrational levels, derived a full set of accurate spectroscopic constants for the $1\nu_7$ and $2\nu_7$ levels, obtained an approximate estimate of the $\nu_7$ bending vibrational frequency, and assigned over 20 previously unidentified lines in the outer envelope of IRC+10216 to vibrationally excited C$_4$H in $1\nu_7$ and $2\nu_7$.

The spectroscopic constants derived from the rotational spectra and the symmetry of the new vibrational levels are well determined, 
but owing to the absence of either supporting  quantum calculations or spectroscopic measurements in other wavelength bands,
the assignments of the rotational spectra to particular vibrational levels were uncertain.
However, with the recent  laser spectroscopic measurements of previously unexplored bending vibrational levels in the ground $X^2\Sigma$ electronic state of C$_4$H \citep{Mazzotti}, we are now able to assign the rotational spectra to four vibrational levels.
The assignments are based in part on the relative intensities of the rotational lines in our discharge source, and theoretical energies 
of the vibrational levels \citep{Graf}, whose accuracy was confirmed in part by the laser spectroscopic measurements \citep{Mazzotti}.


Here we present laboratory measurements of the rotational spectra in four new vibrational levels of C$_4$H in the millimeter band
between 69 and 364~GHz (two each with $^2\Sigma$ and $^2\Pi$ vibronic symmetry), and the two $^2\Sigma$ levels in the
centimeter band between 19 and 38~GHz. 
Also described are interferometric maps with the SMA  of emission from the ground state and $1\nu_7$ level near 257~GHz, which
show for the first time that C$_4$H is present near the central star.
Further interferometric observations at higher angular resolution and sensitivity, might help establish whether other vibrationally excited C$_n$H 
radicals and small carbon bearing reactive species (such as the cumulene carbenes H$_2$C$_n$, etc.) 
are the carriers of some of the many unassigned lines that have been observed in the inner envelope of IRC+10216.
Observations such as these might  in turn help advance the chemistry of the inner envelope of carbon rich AGB stars in the region where 
dust begins to form.



\section{Laboratory and Astronomical Measurements}

\subsection{Laboratory}
\label{subsec:Laboratory}


The C$_4$H radical was observed at millimeter wavelengths with a free space gas discharge spectrometer 
used to study reactive radicals, carbenes, and molecular ions of astrophysical interest \citep{Gottlieb2003}.
Rotational lines in vibrationally excited C$_4$H were measured under conditions which optimized the corresponding transitions 
in the ground vibrational state: a DC~discharge (0.5 A, 1.4~kV) through a flowing mixture of helium (10--15\%) and acetylene (HCCH) 
at a total pressure of 25~mTorr with the walls of the discharge cell cooled to 150~K.
Although the most intense lines of C$_4$H were observed near 300~GHz, measurements with the free space spectrometer 
were extended to frequencies as low as 69~GHz in the three most intense new vibrational levels, and to 44~GHz in $1\nu_7~(^2\Pi)$.
Zeeman chopping with a 50~G magnetic field helped discriminate between paramagnetic species and interfering closed shell species present in the discharge.

\begin{figure}
\begin{center}
\vspace{-1.0cm}
\includegraphics[scale=0.8]{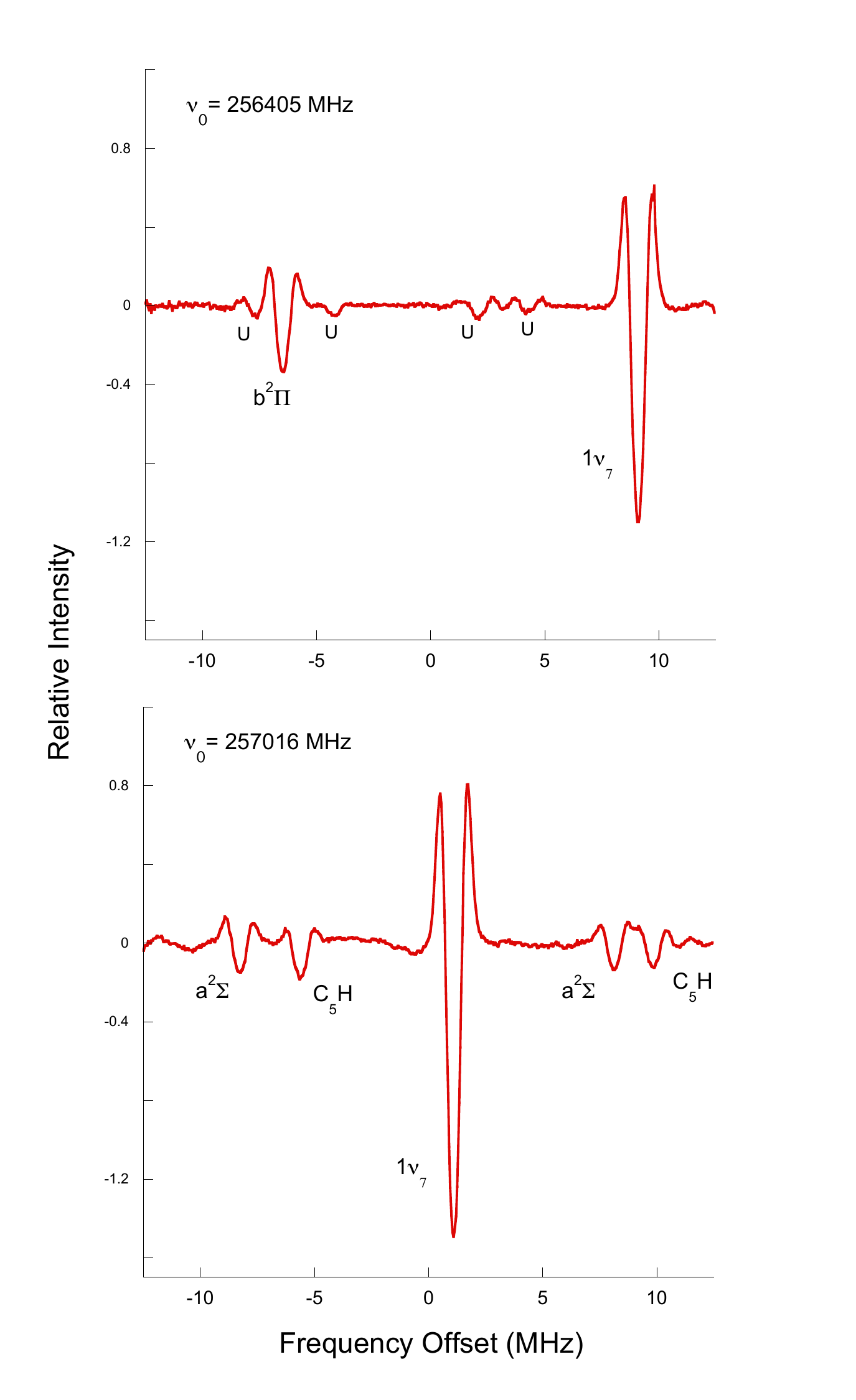}
\vspace{0.0cm}
\caption{\small Sample rotational spectra of the vibrationally excited C$_4$H radical. 
{\it Upper:}  The $J = 26.5 \leftarrow 25.5~^2\Pi_{1/2}, e$ transition in b$^2\Pi$, and the $J=26.5 \leftarrow 25.5~^2\Pi_{1/2}, e$ transition 
in $1\nu_7$  near 256.4~GHz.  The carriers of features labeled U are unidentified.
{\it Lower:} The two spin rotation components of the $N = 27 \leftarrow 26$ transition of a$^2\Sigma$, and the $J=27.5 \leftarrow 26.5~^2\Pi_{3/2}, e$ transition in $1\nu_7$ 
near 257.0~GHz. 
Also present are the two lambda components of the $J=53.5 \leftarrow 52.5~^2\Pi_{3/2}$ transition of the C$_5$H radical.
The line shape is the second derivative of a Lorentzian profile owing to the detection scheme employed. 
The integration time for each spectrum was approximately 15~min.}
\label{fig-MMspectra}
\end{center}
\end{figure}

The frequencies of the rotational lines were derived from a simultaneous least-squares fit of the explicit expression for the second derivative of a theoretical 
Lorentzian profile, and an n$^{\rm{th}}$ order polynomial to spectra observed at 512 points in a 10~MHz bandwidth \citep[see][]{Gottlieb2003}.
The S/N in the raw unsmoothed spectra were  about 10 in approximately 15~min of integration (Figure~\ref{fig-MMspectra}).
Because the line widths were typically about 800~kHz (FWHM),  the second derivative line profile was sampled at approximately 70~points between the two horns (or satellites).
From prior work on similar systems, we estimate that the uncertainties in the measured frequencies are about  20~kHz, or less than 0.1~km~s$^{-1}$ for lines up to 363~GHz.

\begin{figure}
\begin{center}
\vspace{-1.0cm}
\includegraphics[scale=0.9]{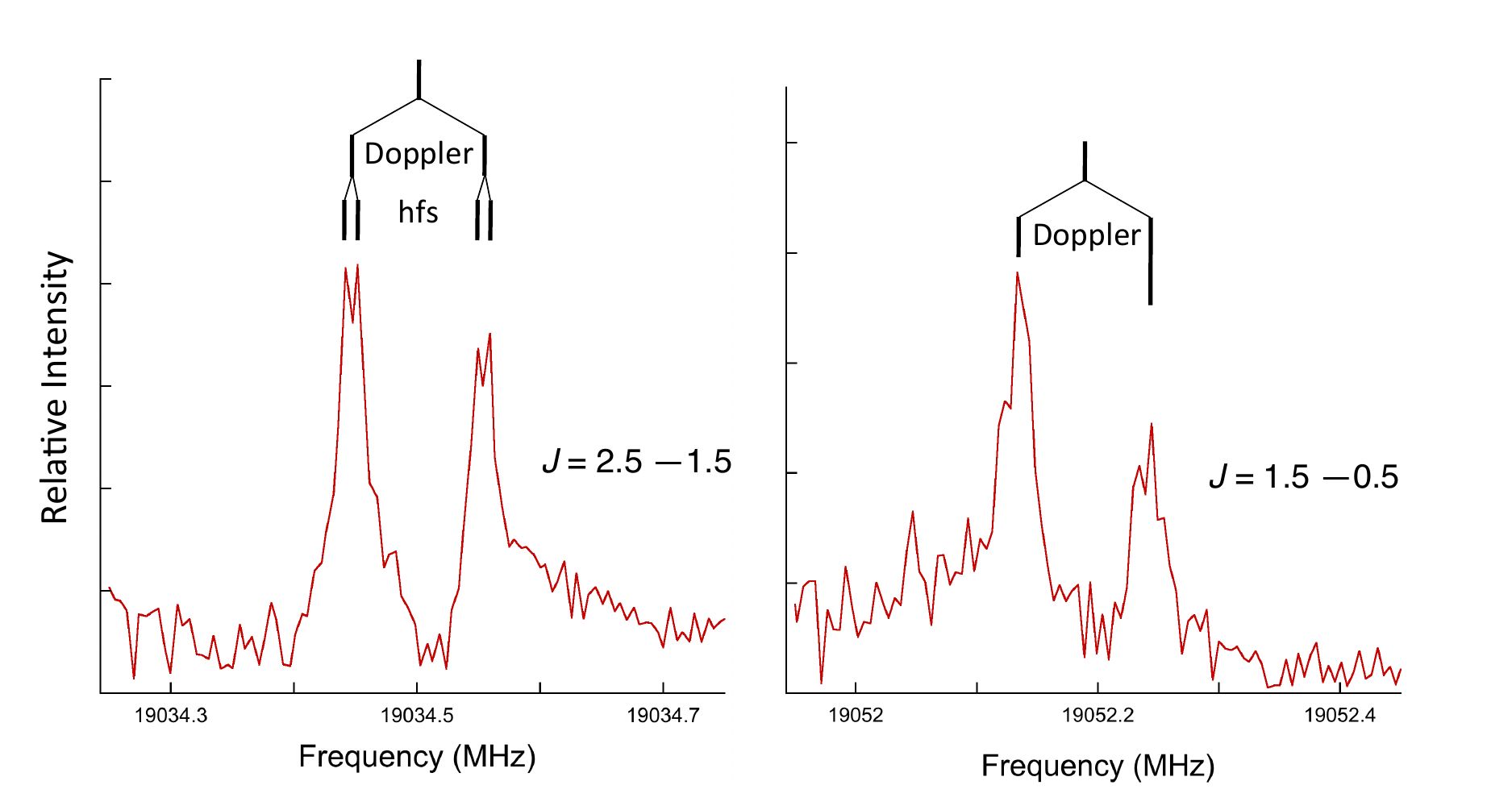}
\vspace{1.0cm}
\caption{Sample rotational spectra in the two spin rotation components of the $N = 2 \rightarrow 1$ rotational transition in the a$^2\Sigma$
excited vibrational level of C$_4$H observed in a supersonic molecular beam.
Partially resolved hyperfine structure is observed in the $J = 2.5-1.5$ component. 
The integration time was 12~min at 19034~MHz and 20~min at 19052~MHz.
}
\label{fig-FTMspectra}
\end{center}
\end{figure}

Three rotational transitions were also observed in the centimeter band with our Fourier transform microwave (FTM) spectrometer
\citep{McCarthy2000} in $2\nu_2$ and the two new ones with $^2\Sigma$ symmetry (see sample spectra in Figure~\ref{fig-FTMspectra}).
The discharge conditions were optimized on the corresponding rotational transition of C$_4$H in the ground vibrational state:
a 1050~V DC discharge in the throat of the supersonic nozzle of the spectrometer
through a mixture of acetylene (0.1\%) heavily diluted in neon. 
Typical gas flow rates were 25~cm$^{-3}$~min$^{-1}$ at standard temperature and pressure (STP), a pulse repetition rate of 6~Hz, 
pulse duration of $300~\mu$s, and stagnation pressure behind the nozzle of 2.5~kTorr. 
Under these conditions, lines in the $1\nu_7$ and $2\nu_7$ levels were  $100-400$ times less intense than those of the ground state.


\subsection{Astronomical}


Spectra and maps of  C$_4$H in IRC+10216 were observed with seven antennas of the SMA in the compact configuration on 6~March 2008, 
at baselines between 16.4 and 77.0~m.
The synthesized beam was $2.92^{\prime\prime} \times 2.60^{\prime\prime}$ at a position angle (P.A.) of $72^{\circ}$, and the
field of view (size of primary beam) was $\sim 46^{\prime\prime}$.
The system temperature ranged from 200 to 350~K, $\tau_{\rm{225~GHz}} \sim 0.20$, and the total on source integration time was about 10~hours.
The correlator was configured for a uniform frequency resolution of 0.64~MHz per channel in the lower and upper 2~GHz wide sidebands 
with band centers at 256.4 and 266.4~GHz.
The quasar 0854+201 was observed every 20~min for gain calibration.
The spectral bandpass was calibrated by observations of 3C273 and Jupiter, and absolute flux calibration by observations of Mars and Titan.




\section{Results}
\label{sec:results}

\subsection{Laboratory Spectroscopic}

\label{sec:LabSpectroscopic}

\begin{figure}
\begin{center}
\vspace{0.0cm}
\includegraphics[scale=0.6]{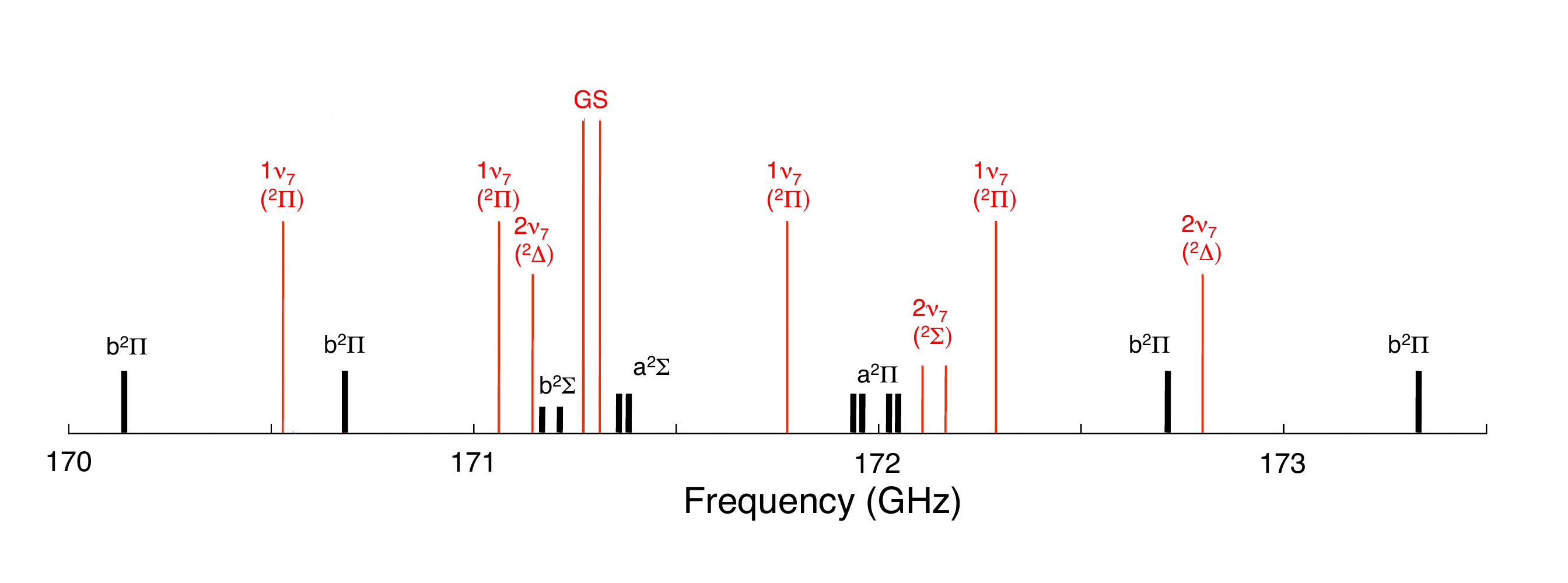}
\vspace{2.0cm}
\caption{Stick spectrum of the $N = 18 \leftarrow 17$ rotational transition in the ground and excited vibrational 
levels of C$_4$H near 171~GHz. 
{\it Red:} Three vibrationally excited levels identified previously by \citet{Yamamoto}, and the $^2\Sigma$ electronic ground state ground state (GS).
{\it Black:} The four new vibrational levels a$^2\Pi$, b$^2\Pi$, a$^2\Sigma$, and b$^2\Sigma$ (this work).  
The relative intensities were measured in a low pressure discharge through helium and acetylene (see Section~\ref{subsec:Laboratory}).}
\label{fig-scheme}
\end{center}
\end{figure}

Four new series of satellite lines were assigned  to rotational transitions in vibrationally excited C$_4$H because:  
(1) the rotational and centrifugal distortion constants ($B$ and $D$) are within 0.4\% of those in the ground state 
(see Figure~\ref{fig-scheme}); and
(2) the analogous satellite lines were observed in C$_4$D, when HCCH was replaced by DCCD in our discharge. 
The symmetries of the four new vibrational levels were established from the following considerations.  
Two quartets, symmetrically displaced about a central frequency, were assigned to two new $^2\Pi$ levels
(designated $a^2\Pi$ and $b^2\Pi$). 
The quartets, consisting of four lines with comparable intensities in each rotational transition, precluded levels with $^2\Sigma$ symmetry 
and the well-resolved $K$-doubling implied that $K = 1$.
The remaining two series were assigned to $^2\Sigma$ levels ($a^2\Sigma$ and $b^2\Sigma$), 
because only two lines were found in each series and the coupling case remained Hund's case ($b$) at all observed frequencies.
Although doublets (rather than quartets) are frequently observed in levels with $K > 1$  because the $K$-doubling is 
often unresolved (e.g., in $^2\Delta$ levels), there is a significant departure from case ($b$) coupling in the lower rotational 
transitions in vibrational levels with $K > 1$ owing to the spin-orbit interaction.
Listed in Tables~\ref{tab-nu7pi}--\ref{tab-sib} are the measured rotational lines in the four new vibrational levels, and some new 
measurements in the previously studied $1v_7$ ($^2\Pi$) and $2v_7$ ($^2\Sigma$ and $^2\Delta$) levels.



In a closed shell molecule, $l$-type doubling is described by only one constant ($q$), but two constants ($p$ and $q$) 
are needed to describe $K$-type doubling in molecules with vibrational and electronic contributions \citep{brown}. 
In the well studied C$_2$H radical, the $A^2\Pi$ excited electronic state is 3600~cm$^{-1}$ above the $X ^2\Sigma$ ground  
state and $p$ is much smaller than $q$, implying that the vibrational contribution to the $K$-type doubling is dominant 
\citep{Killian,Sharp}.  
However, $p$ and $q$ are comparable in C$_4$H, because  the $A^2\Pi$ state is much closer to ground  
\citep[222~cm$^{-1}$;][]{Mazzotti}, and the vibronic interaction between the two electronic states in C$_4$H is much greater than 
in C$_2$H.


The rotational spectra in the vibrationally excited levels of C$_4$H were analyzed by essentially the same method 
as previously applied to vibrationally excited C$_2$H  \citep[and references therein]{Killian}.
In this approach, the rotational structure in each vibrational level was described by an effective Hamiltonian 
used to analyze an isolated electronic state in a diatomic molecule.
In open shell molecules, the electron orbital ($\Lambda$) and vibrational ($l$) angular momentum are coupled, and
$\Lambda$-type and $l$-type doubling combine to produce $K$-type doubling.
It has been shown in work similar to that here that the rotational spectra in excited vibrational levels of open shell molecules
are reproduced to very high accuracy by this approach.
Some rotational levels in $b^2\Sigma$ and $b^2\Pi$ were perturbed by nearby vibrational levels, requiring two additional terms 
in the Hamiltonian to reproduce the spectrum of  $b^2\Sigma$ (see Section~\ref{sec:perturbations}).


We have adhered to the sign convention in \citet{Killian}, in which the sign of $K$-doubling constant $q$ was
determined on the assumption that the hyperfine doubling constant $d$ is positive.
Following this convention, $q$ is negative in $1\nu_7~(^2\Pi)$.  
Although $d$ is not well determined in the $b^2\Pi$ level, $q$  is assumed to be negative in $b^2\Pi$ as well.
Because the hyperfine constants were not determined in the $a^2\Pi$ level, we assumed by analogy with $1\nu_7~(^2\Pi)$ and $b^2\Pi$
that  $p$ is also negative in the $a^2\Pi$ level.
The signs of $\gamma$ and the hyperfine coupling constants $b_{\rm{F}}$ and $c$ are the same as those in the ground state as expected, because the spin-rotation and hyperfine interactions are sufficiently decoupled from vibrational effects.
The spin-orbit constant $A$ is negative in the $1\nu_7~(^2\Pi)$, $2\nu_7~(^2\Delta)$, $a^2\Pi$, and $b^2\Pi$ vibrational levels of  the 
$X^2\Sigma$ ground electronic state, because $A$ is negative in the low-lying $A^2\Pi$ excited electronic state \citep{Mazzotti}.


\subsubsection{Perturbations}
\label{sec:perturbations}

The $b^2\Sigma$ and $b^2\Pi$ levels exhibit pronounced $N$-dependent perturbations by vibrational levels that 
were not observed in this study.
In $b^2\Sigma$, the interacting levels cross between $N = 15$ and 25, resulting in very large shifts in the transition 
frequencies of one of the spin-rotation components, whereas the other component is not perturbed.  
To fit this pattern, a standard Coriolis-type contribution,
\begin{equation}
\Delta E ~=~ {{q_l (N^2-1) }\over{1 ~+~ \rho N(N+1) }}, \label{pert}
\end{equation}
was added to the energies of the perturbed component.  
The $N$-dependence of the numerator depends on the specific spin-rotation components in the interaction. 
We chose the pair that gave the lowest standard deviation of the fit, but $q_l$ differs by $\le10$\% for other choices and the constants 
differ by less than one standard deviation.
In $b^2\Pi$, all four components are weakly perturbed and the magnitude of the perturbation increases with increasing $N$.  
Analyzing the perturbation in $b^2\Pi$ would require adding several empirical parameters with no clear 
interpretation.
Instead, the two most strongly perturbed transitions in $b^2\Pi$ were removed from the fit, and $b^2\Pi$ was treated as though it
were an isolated $^2\Pi$ state (see Table~\ref{tab-pib}).


\subsubsubsection{$b^2\Sigma$}
\label{subsec:bSigma}


The strong perturbation of the $N=J+\frac{1}{2}$ component of the spin-rotation doublets in $b^2\Sigma$ implies that 
there is an interaction with a level with either $^2\Pi$ or $^2\Sigma^-$ vibronic symmetry.  
To first order in perturbation theory, $q_l =  \beta^2/\Delta\nu$ and $\rho = \Delta B/\Delta\nu$
in Eq.~(1), where
$\Delta\nu$ is the difference in band origins $\nu_0(b^2\Sigma) - \nu_0(x)$,
$\Delta B = B(b^2\Sigma) - B(x)$, 
 the perturbing level is labeled $x$,
and $\beta$ is the coefficient of the Coriolis coupling matrix element. 
Both $q_l$ and $\rho$ in $b^2\Sigma$ are negative (Table~\ref{NewConst}), indicating that the perturbing level has a smaller 
$B$ and the  band origin is higher than $b^2\Sigma$.

\subsubsubsection{$b^2\Pi$}



The $b^2\Pi$ level is perturbed at high $N$ by a level that has not yet been identified.
When the two uppermost rotational lines from any one of the four parity/spin-orbit branches are removed from the fit,
the rms is reduced by a factor of two and the higher order distortion terms $L$ and $p_H$ are not needed.
In Table~\ref{NewConst}, 
transitions with $N' > 20$ in $J~=~N-\fr{1}{2}~e$  were omitted because doing so yielded the lowest rms,
but other fits with comparable rms were obtained if one of the other branches was neglected instead.  
The addition of higher-order terms in the Hamiltonian such as $q_H$, implies that more than one branch is significantly perturbed.
A perturbation analysis similar to that for $b^2\Sigma$ was not attempted for $b^2\Pi$, primarily because 
the frequency shifts are much smaller in $b^2\Pi$.  
In the future, measurements of the perturbing levels near the resonance interactions in $b^2\Sigma$ and $b^2\Pi$
might yield accurate coupling constants and frequency separations of the perturbing levels.

 
\subsubsection{Assignment of Vibrational Modes}

The vibrational modes of C$_4$H may be loosely divided into C--H local modes ($\nu_1$ stretch and $\nu_5$ bend), 
and C$_4$ normal modes ($\nu_2$ and $\nu_4$ symmetric stretches, $\nu_3$ antisymmetric stretch, and $\nu_6$ and $\nu_7$ bends). 
For the most likely ones observed here ($\nu_4, \nu_5, \nu_6, \nu_7$, and combinations thereof), 
accurate estimates of the band origins of three ($\nu_4$, $\nu_7$, and $\nu_6 + \nu_7$) were recently derived from 
laser spectroscopic double resonance four-wave mixing measurements of the $B^2\Pi - X^2\Sigma^+$ electronic band near 
24,000~cm$^{-1}$ \citep{Mazzotti}.
Theoretical frequencies of all seven vibrational modes were calculated independently in the harmonic approximation \citep{Graf}.
Because the theoretical frequencies of $\nu_4$ and $\nu_7$ agree to within 6\% or better with those measured by Mazzotti et al., 
the theoretical frequencies of $ \nu_5$ and  $\nu_6$ are likely known to comparable accuracy (see Table~\ref{tab-hc3n}). 

Owing to the precision of the frequency measurements and degree of symmetry in the rotational spectra here,
the vibronic symmetries and spectroscopic constants of the four new levels in C$_4$H were determined to a very high level of confidence.
Because the rotational information needed to guide the assignment of the new vibrational levels of C$_4$H was not available either from 
theoretical calculations or IR spectroscopic measurements, the assignments to particular  vibrational levels is tentative.


Rotational spectra in vibrational satellites of closed shell polyatomic molecules are routinely assigned on the basis 
of relative intensities of the lines, vibronic symmetry, the magnitude and sign of the vibration-rotation interaction 
$\alpha$ (where $B_v - B_0 = - v\alpha$), and the $l$-doubling constant $q$, where the approximations
\begin{equation}
\alpha (v_1,v_2,v_3,...) ~=~ v_1 \alpha_1 ~+~ v_2 \alpha_2
~+~ v_3 \alpha_3 + ... \label{alpha}
\end{equation}
and
\begin{equation}
q_n ~=~ 2.6 \frac{B^2}{\nu_n} \label{qn}
\end{equation}
are normally accurate to within 20\% for low vibrational excitation.
For example, in the well studied linear cyanopolyynes HCN and HCCCN  \citep{Mbosei}, the magnitude of $\alpha$ increases as the 
fundamental frequency of the mode decreases, it is negative in bending levels (${B_v}^{\rm{bend}} > B_0$), and positive in stretching levels 
(${B_v}^{\rm{stretch}} < B_0$) as expected.
Unfortunately, the standard procedure for assigning rotational spectra to specific vibrational levels in closed shell polyatomic molecules 
is not necessarily applicable in the open shell C$_n$H radicals.
For example, $\alpha$ in the $1\nu_7$ level of C$_4$H is 1/5 that in $2v_7$, rather than 1/2 as expected from Eq.~\ref{alpha}.
Similarly, the $\alpha$'s in the bending vibrational levels in C$_2$H are anomalous:  $\alpha$ in the $1\nu_2$ level 
is 1/12 that in $2\nu_2$, rather than 1/2 as expected \citep{Killian}.


Assignments of the new vibrationally excited levels of C$_4$H were restricted to levels with   $^2\Sigma$ and $^2\Pi$ symmetry, 
and to ones whose magnitude of $\alpha$ are among the smallest.
Before considering higher excitations with $v ~\geq ~2$ and higher-lying combination levels, we examined 
the fundamental modes ($\nu_4, \nu_5, \nu_6,~{\rm{and}}~ \nu_7$) and combinations with the $\nu_7$ bend. 
The requirement that the rotational frequencies in the new vibrational levels are close to those of the ground state,
implies that they are Hund's case ($b$) coupled in the frequency range where most of the measurements were made (i.e., at $N \approx 30$).  
Because the spin-orbit interaction is small, the vibration-rotation coupling increases with increasing vibrational excitation 
\citep[as indicated by Eq.~\ref{alpha}, and observed in HCCCN;][]{Mbosei}, implying that rotational lines in the lower 
vibrational levels lie in the frequency band that we had covered.  


\subsubsection{Spectroscopic considerations}
\label{subsec:spectroscopic}

There are many vibrational levels  in the ground $X^2\Sigma$ and excited $A^2\Pi$ electronic states of C$_4$H below 
1000~cm$^{-1}$  \citep[see Figure~6 in][]{Mazzotti}.
Considering only those in the  $X^2\Sigma$ electronic state, (excluding $1\nu_7$ and $2\nu_7$) there are eight possible candidates.\footnote{The eight candidates in the  $X^2\Sigma$ electronic state are 
 ($\nu_4, \nu_5, \nu_6, \nu_7) = (0003)$, (0004), (0010), (0100), (1000), (0011), (0101), and (1001).}
Three of these, all with excitation energies between 400 and 600~cm$^{-1}$, were considered as candidates for the two new vibrational levels with $\Pi$ symmetry:
$1\nu_6$, $1\nu_5$, and $3\nu_7$.  
We have assumed on the basis of the qualitative form of Eq.~\ref{alpha}, that $\alpha$  in $3\nu_7$ is significantly larger than in $2\nu_7$. 
Therefore the $3\nu_7$ level was ruled out, because the rotational lines in the levels we observe are closer in frequency to the ground state lines than those in $2\nu_7$.
As a result, the $b^2\Pi$ level was assigned to  $1\nu_6$, and the less intense $a^2\Pi$ level to the higher lying $1\nu_5$ level.  


There are seven candidates (between 600 and 1000~cm$^{-1}$ above ground) for the two new levels with 
$\Sigma$ symmetry.
The $4\nu_7$ level was discounted, owing to the large $\alpha$ and intensity considerations (see Section~\ref{subsec:intensity}).
The $a^2\Sigma$ level was assigned to the $\nu_6 + \nu_7$ combination level, 
even though the magnitude of  $\alpha$ is very small for a combination level and the $^2\Delta$ and the second $^2\Sigma$ level 
were not observed, because $\nu_6 + \nu_7$ has the lowest energy of the remaining levels with $\Sigma$ symmetry. 
The  $b^2\Sigma$ level was assigned last, because it was made primarily on the basis of the intensity information discussed in 
Section~\ref{subsec:intensity}.


\subsubsection{Intensity considerations}
\label{subsec:intensity}

Assignments of the vibrational levels of C$_4$H on the basis of relative intensities might be secure if  the vibrational temperature
($T_{\rm{vib}}$) in our HCCH discharge were known precisely. 
Our best independent estimate of $T_{\rm{vib}}$ is from prior work on the C$_2$H radical because:
(1) the electronic and vibrational level structures of C$_2$H and C$_4$H are qualitatively similar; 
(2) both radicals are observed under comparable excitation conditions in the same HCCH discharge; and 
(3) there is a large body of IR measurements and extensive quantum calculations of the vibrational level structure in C$_2$H 
\citep[and references therein]{Sharp}.
We found that $T_{\rm{vib}}$ in the  bending vibration (010), and the  combination levels (110) and (011) of C$_2$H
is $100-200$~K higher than the temperature of the wall of our discharge cell (150~K), whereas $T_{\rm{vib}}$ 
in the stretching levels (100) and (001) is about 5~times higher \citep[$\sim 1500$~K;][]{Killian}.
By analogy with C$_2$H, it was anticipated that $T_{\rm{vib}} = 250 - 350$~K would also apply to C$_4$H in $\nu_5, \nu_6, \nu_7$, 
and in levels in combination with the bending vibrations.
As a test, we first examined the excitation of $1\nu_7$ and $2\nu_7$ in C$_4$H, because the assignments and the vibrational energies 
are well established \citep{Yamamoto,Mazzotti}.
The derived vibrational temperature for the $1\nu_7$ and $2\nu_7$ levels in C$_4$H ($T_{\rm{vib}} = 311 \pm 66$~K) 
is in good agreement with that of C$_2$H ($370 \pm 120$~K) and C$_2$D \citep[$260 \pm 60$~K;][]{Killian}, therefore
the analogy with C$_2$H appears justified.

\begin{figure}
\begin{center}
\vspace{-1.5cm}
\hspace{-0.25cm}
\includegraphics[scale=0.85]{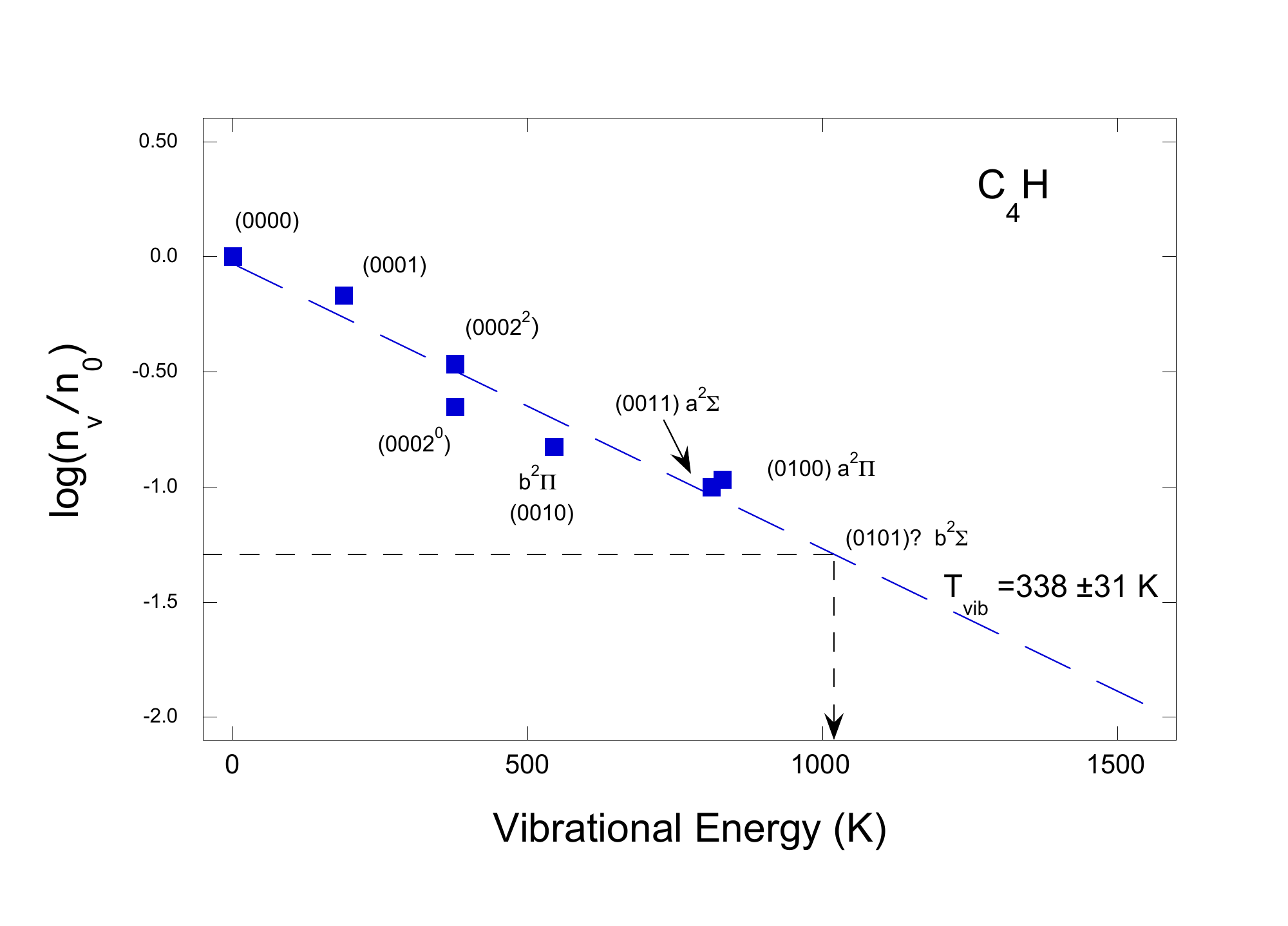}
\vspace{-1.0cm}
\caption{Vibrational temperature diagram of the $1\nu_7$, $2\nu_7$, and four new levels of  C$_4$H in the $X^2\Sigma$ electronic state.
The relative populations in the ground ($n_0$) and excited vibrational levels  ($n_{\rm{v}}$) were determined from measurements of the  
$N = 15 \leftarrow 14$ rotational transition near 143~GHz in a low pressure discharge through He and HCCH cooled to 150~K.
The vibrational levels are labeled as $(\nu_4,\nu_5,\nu_6,\nu_7)$.
See Sections~\ref{subsec:spectroscopic} and \ref{subsec:intensity} for a discussion of the assignments of the new levels.} 
\label{VTdiag}
\end{center}
\end{figure}

Encouraged by the similarity between the vibrational excitation of C$_2$H and the lowest bending vibrational levels of C$_4$H, 
we then  considered the intensity information when assessing the assignments of the new levels that have been made
on the basis of the spectroscopic constants (Section~\ref{subsec:spectroscopic}).
The $1\nu_4$ level at 930~cm$^{-1}$ \citep[1337~K;][]{Mazzotti} was ruled out because $T_{\rm{vib}} \sim 450$~K is much too low for a stretching vibration, and 
the intensity is much too high if $b^2\Sigma$ were a combination level with a bending vibration (e.g., $\nu_4 + \nu_7$ at 1579~K above ground). 
On the assumption that $T_{\rm{vib}}$ is comparable to that of the $1\nu_7$ and $2\nu_7$ bending vibrational levels, the most plausible assignment of $b^2\Sigma$ 
is the combination level $\nu_5 + \nu_7$.
The vibrational energy of $b^2\Sigma$ derived from  Figure~\ref{VTdiag} ($1015 \pm 98$~K, or $706 \pm 68$~cm$^{-1}$) is in good 
agreement with the theoretical frequency \citep[756~cm$^{-1}$;][]{Graf}.
With these assignments, the relative intensities of all six vibrational levels of C$_4$H observed in our discharge 
are characterized by a vibrational temperature ($338 \pm 31$~K; Figure~\ref{VTdiag}) that is consistent with the bending vibrational levels of 
C$_2$H.


\subsection{Astronomical Observations}

\begin{figure}
\begin{center}
\vspace{-1.0cm}
\hspace{-0.5cm}
\includegraphics[scale=0.75]{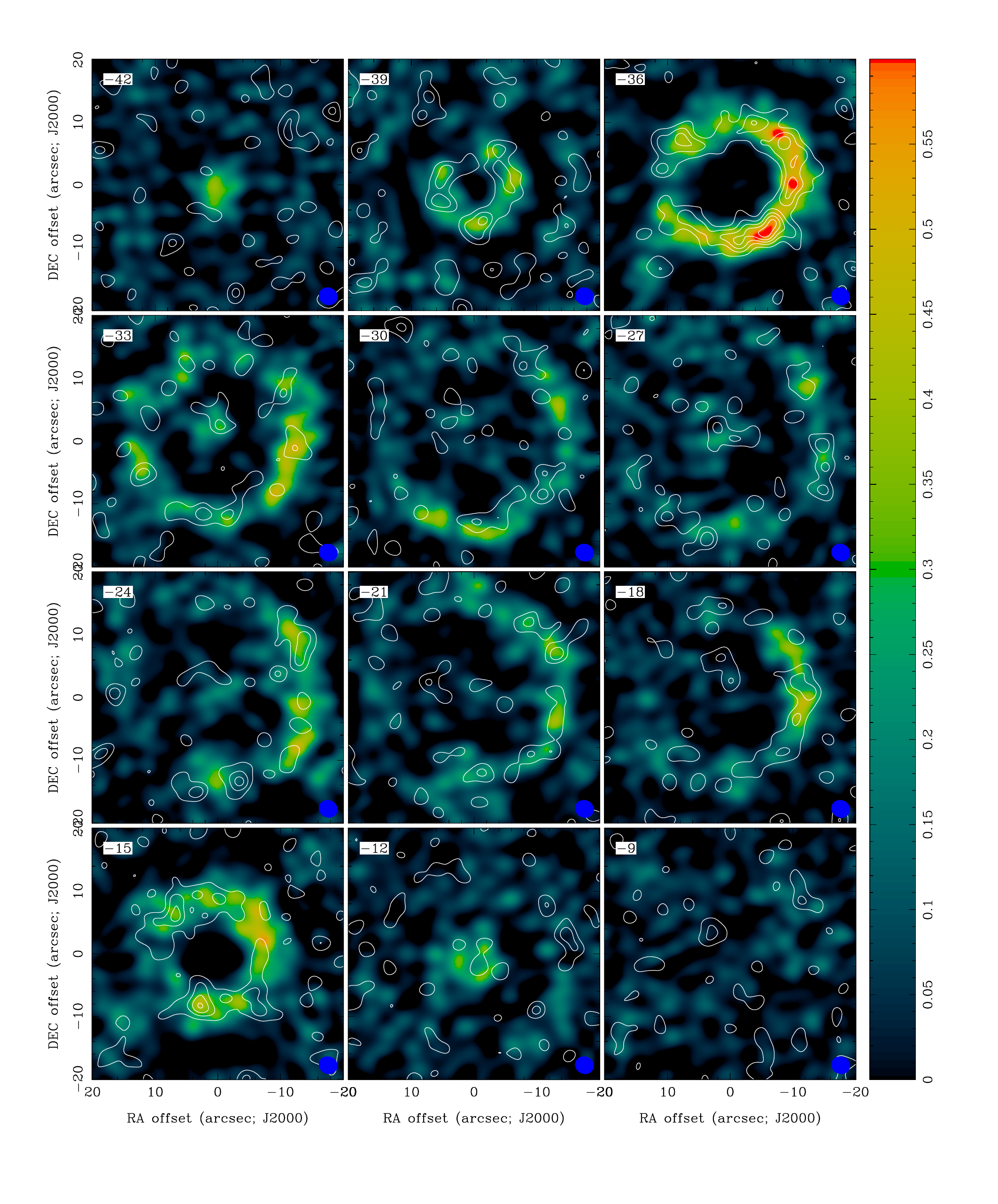}
\vspace{-1.0cm}
\caption{\small Velocity channel maps of the $N = 27 \rightarrow 26$ rotational transition of C$_4$H in the ground and $1\nu_7~(^2\Pi)$
excited vibrational level. 
{\it Color half-tone:} average of the two spin doublets in the ground vibrational state at 256880 and 256919~MHz at $E_u = 173$~K.
{\it White contours:}  average of three of the four lambda components of the $1\nu_7~(^2\Pi)$ level at 256414, 257017, and 257218~MHz
at $E_u = 363$~K. 
The contours are plotted every $3 \sigma$, starting with $3 \sigma$ and an rms noise of 20~mJy/beam.
The actual rms noise varies in each channel map, but is approximately 20 to 50~mJy/beam.
The maps show the close correspondence between the ground and excited state emission, and the  azimuthal asymmetry and clumpiness in the expanding 
outer shell.
}
\label{ChanMaps}
\end{center}
\end{figure}

\begin{figure}
\begin{center}
\vspace{-3.0cm}
\includegraphics[scale=0.8]{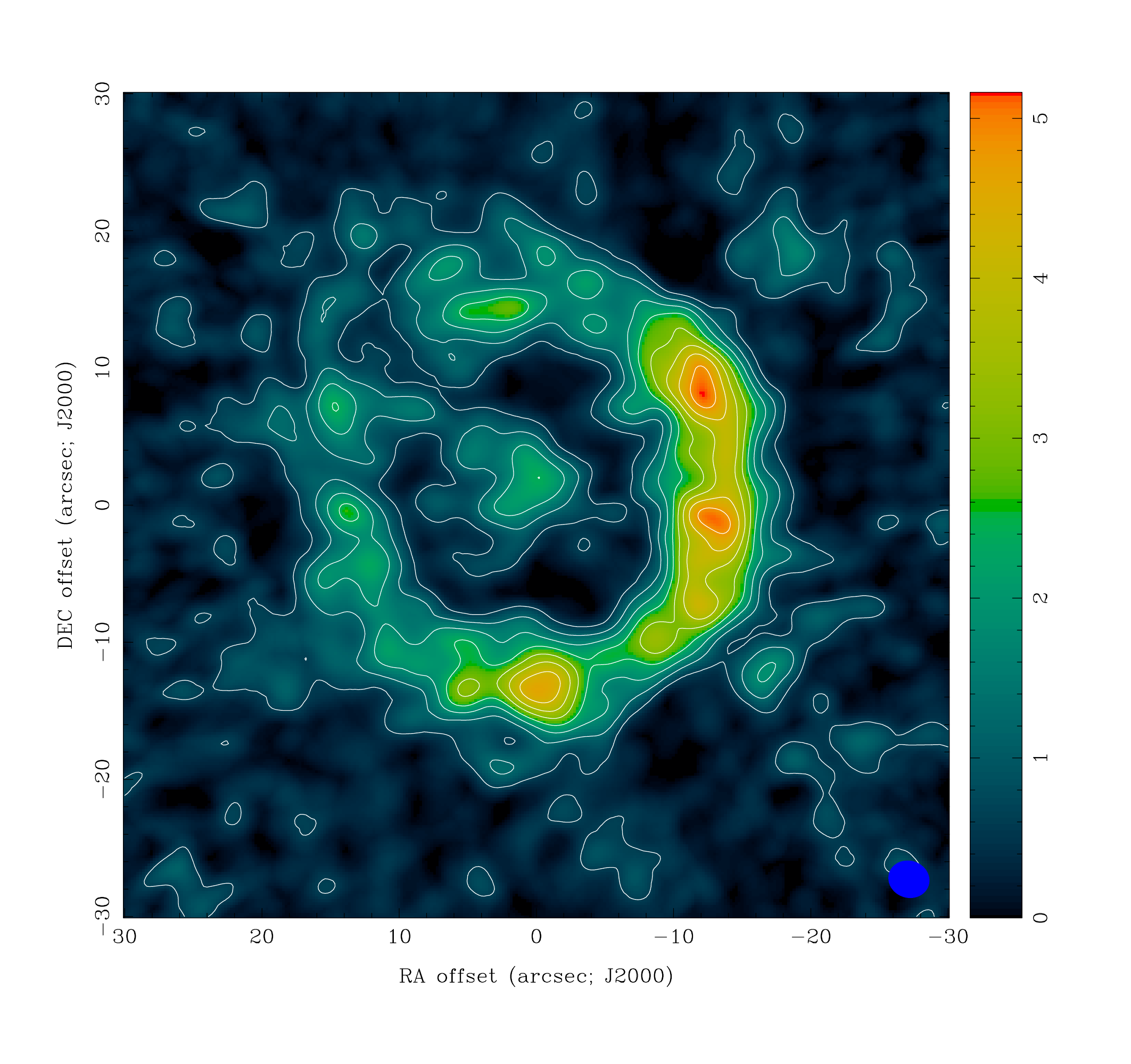}
\vspace{-0.5cm}
\caption{\small Integrated intensity map of the ground state emission of C$_4$H (see Figure~\ref{ChanMaps}).
The velocity range of integration ($-33~{\rm{to}}~ -18$~km~s$^{-1}$) is approximately centered on the systemic velocity.
The emission near the center shows that C$_4$H is not confined solely to the outer envelope.
}
\label{v0mom0}
\end{center}
\end{figure}

\begin{figure}
\begin{center}
\vspace{-3.0cm}
\includegraphics[scale=0.8]{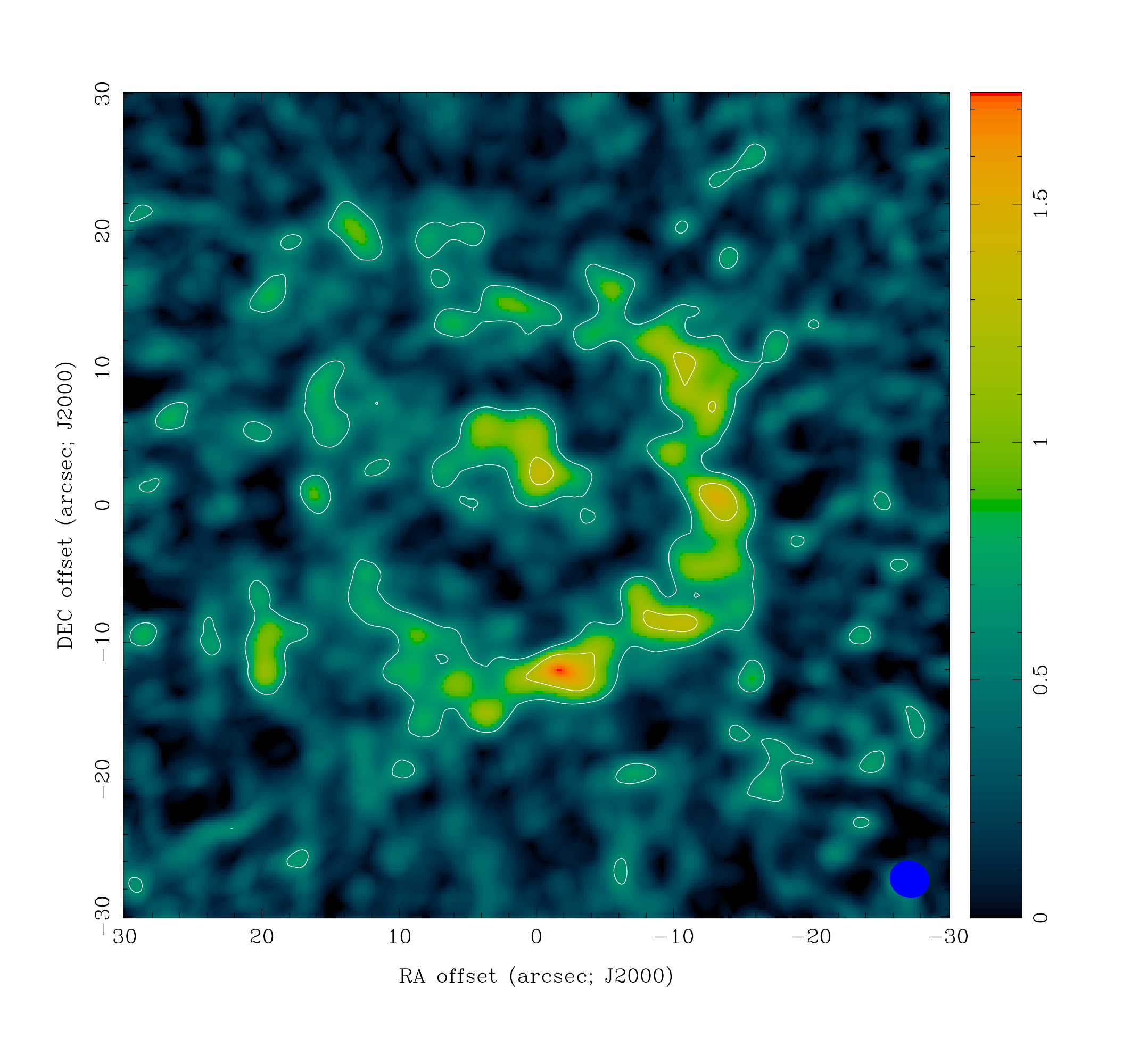}
\vspace{-0.5cm}
\caption{\small Integrated intensity map of C$_4$H of the $1\nu_7$ excited vibrational level (see Figure~\ref{ChanMaps}). 
The velocity range of integration ($-33~{\rm{to}}~ -18$~km~s$^{-1}$) is approximately centered on the systemic velocity.
The map here of the integrated intensity of the $1\nu_7$ level confirms that  C$_4$H is present close to the star.
The clumps seen in the outer ring correlate well with those seen in the ground state emission in Figure~\ref{v0mom0}.
}
\label{v1mom0}
\end{center}
\end{figure}

Shown in Figure~\ref{ChanMaps}, are channel maps of  the $N = 27 \rightarrow 26$ rotational transition of C$_4$H near 257~GHz in the ground vibrational state, and the $1\nu_7~(^2\Pi)$ excited vibrational level. 
The rotational transitions observed here are about 2.5 times higher in frequency, and 150~K higher in energy ($E_u$) than those observed earlier in the
3~mm band with the IRAM Plateau de Bure Interrferometer (PdBI) by \citet{Guelin1999}.
 The peak flux density is 0.63~Jy/beam for the average of both spin components in the ground state, and 0.34~Jy/beam for three of the four rotational transitions in the vibrationally excited level.
Also shown are integrated intensity maps of  the same transitions in the ground state and excited vibrational level,
for a velocity interval  corresponding to the middle two rows in the channel maps (from $-33~{\rm{and}} -18$~km~s$^{-1}$ in Figure~\ref{ChanMaps}) that is roughly centered on the systemic velocity of $-26$~km~s$^{-1}$ (see Figures~\ref{v0mom0} and \ref{v1mom0}).


As noted earlier by \citet{Guelin1999}, the ground and excited emission of C$_4$H follow each other very closely in the outer envelope.
Here for the first time, both ground and vibrationally excited C$_4$H are observed near the central star.
The exact nature of the emission near the central star remains unclear from our SMA observations --- i.e., whether it is from a smaller expanding inner
shell that has not reached the terminal velocity ($\sim 15$~km~s$^{-1}$), or an asymmetrical structure connecting the outer shell to the inner envelope. 
Interpretation of the central emission will require further interferometric observations at  higher S/N and with a much better $u$-$v$ coverage.
 The clumpy features in the expanding shell structure illustrates the inherent uncertainties when comparing the interferometric measurements  with published chemical models (which typically show only radial dependence). 
As expected, the flux density in the ground and vibrationally excited emission in the central region (0.6 versus 0.4~Jy/beam) is comparable at the higher kinetic temperature of the gas \citep{Fonfria2008}.


We also examined the 345~GHz spectral line survey with the SMA \citep{Patel2011} for possible evidence of the four new excited bending vibrational levels of C$_4$H, but none was found at peak flux densities averaged over a $3^{\prime\prime}$ synthesized beam of $\ge 0.15 $~Jy 
in the $294-296$ and $304-306$~GHz range.


\section{Discussion}
\label{sec:discussion}


It has generally been assumed that C$_4$H (and other C$_n$H radicals) is confined to the outer envelope of IRC+10216, 
because that is what is implied from the spectral line profiles of the ground and $1\nu_7$, $2\nu_7~(^2\Sigma)$, and $2\nu_7~(^2\Delta)$ excited vibrational levels observed at low angular resolution with single antennas. 
Furthermore, interferometric maps of rotational transitions of C$_4$H in the ground vibrational and the lowest excited vibrational level $1\nu_7$
in the 3~mm band, show that the  emission is confined to a thin ($3^{\prime\prime}$ thick) shell in the outer envelope with a radius ($r $) of $15^{\prime\prime}$ \citep{Guelin1999} --- just as most chemical models predict.
However, \citet{Cordiner2009} recently showed that when density-enhanced dust shells are included in the chemical models,
C$_4$H is predicted to be present in a thick inner ring ($r  \sim 8^{\prime\prime}$) in addition to the outer ring at $15^{\prime\prime}$. 


Rotational lines in the four new vibrational levels of C$_4$H will most likely be observed in the inner envelope of IRC+10216, because that is where these higher lying levels will be most populated owing to the higher densities and temperatures closer to the star.
Although we did not find any evidence for lines of the new excited bending vibrational levels of C$_4$H, it is plausible that the new vibrationally excited levels of C$_4$H will be observed when millimeter-wave interferometers attain higher sensitivity now that we have established that C$_4$H is present near the central star.


Direct observations of HCCH in the IR and HNC in the millimeter band, provide a template for assigning and analyzing rotational spectra in vibrationally excited levels 
of C$_n$H radicals among the many unassigned features recently observed in the inner envelope of IRC+10216 with interferometers.
These include: 
(1) a detailed analysis of  the kinetic, vibrational, and rotational temperatures of HCCH at  $1R_{\star} \le r \le 300R_{\star}$ derived from rotationally resolved vibrational lines  in the $11 - 14~\mu$m band \citep{Fonfria2008};  
(2) observations  of HNC in the dust formation zone near 270~GHz with ALMA Cycle~0, and an accompanying detailed radiative transfer analysis \citep{Cerni2013}; and 
(3) a related observation of HCCCN in the dense core of the protoplanetary nebula CRL~618 with the SMA in the 345~GHz band 
at  $r < 0.7^{\prime\prime}$ from the central star where $T  = 400 - 600$~K  \citep{Lee2013}.


Sample spectra obtained in ALMA Cycle~0 with sixteen 16~m diam antennas and a $0.6^{\prime\prime}$ synthesized beam in the $263-272$~GHz band  \citep{Cerni2013} are about 7 times more sensitive than those in the SMA survey in the 345~GHz band 
\citep{Patel2011}.\footnote{\footnotesize The collecting area of ALMA Cycle~0 was 14~times greater, but the integration time (with overhead) was about $3-4$ times greater with the SMA.  The synthesized beam in ALMA Cycle~0 was 5 times smaller, but the diameter of the inner envelope 
($0.2^{\prime\prime}$) is small with respect to the angular resolution in both observations.}  
Referring to Figure~1 in \citet{Cerni2013}, the peak flux densities of most of the unassigned narrow lines observed with ALMA ($\le 0.15$~Jy/beam)  are below the detection limit with the SMA.
When ALMA attains full design sensitivity with 66~antennas it will be about 4 times more sensitive still, offering great promise for the identification of key reactive molecular intermediates such as the C$_n$H radicals in the inner envelope of the prototypical carbon rich 
AGB star IRC+10216, and elucidation of the chemical processes crucial to the formation of carbonaceous dust in the interstellar gas.



\onecolumn

\newpage

\begin{deluxetable}{llc ccc cc }
\tablecaption{Vibrationally Excited C$_n$H Radicals in IRC+10216 }
\tabletypesize{\normalsize}
\tablewidth{0pt}
\tablehead{
\multicolumn{1}{l}{Molecule} 
&\multicolumn{1}{l}{Level\tablenotemark{a}}
& \multicolumn{1}{c}{Band Origin\tablenotemark{b}} 
&\multicolumn{1}{c}{}
&\multicolumn{1}{c}{Beamwidth}
&\multicolumn{1}{l}{Flux\tablenotemark{c}}
&\multicolumn{2}{l}{Reference}   \\

\cline{7-8}

\multicolumn{1}{c}{}  
&   \multicolumn{1}{c}{}  
&   \multicolumn{1}{c}{ (cm$^{-1}$)}  
&  \multicolumn{1}{c}{}
&   \multicolumn{1}{c}{($^{\prime\prime}$)}    
&  \multicolumn{1}{l}{(Jy)}
&  \multicolumn{1}{l}{Space}
&  \multicolumn{1}{l}{Lab}

 }
\startdata

C$_2$H                      & $1\nu_2~(^2\Pi)$                     &   370       &         &       29          &       0.3                   &  d       &  e    \\

C$_3$H                      &$1\nu_4~(^2\Sigma^u$)          &   27         &         &   28              &    0.4                       & d         &  f      \\

C$_4$H                      &$1\nu_7~(^2\Pi)$                      &   170       &         &     27            &             2.9             &  d        &  g          \\
                                     &$2\nu_7~(^2\Sigma)$              &   375       &         &     27            &             1.5             &  d        &  g        \\
                                     &$2\nu_7~(^2\Delta)$                &   375       &         &     27            &             0.6             &  d        &   g         \\

C$_6$H                      &$\nu_{11}?~(^2\Sigma)$         &     14         &         &       28          &     0.1                      &  h        &  i    \\
                                     &$\nu_{11}?~(^2\Delta)$           &     73         &         &      28           &    0.04                    &  h        &  i        \\

\enddata

\tablenotetext{a}{Vibrational levels whose rotational spectra have been observed in IRC+10216.  } 
\tablenotetext{b}{Approximate energy of the band origin above ground.} 
\tablenotetext{c}{Peak flux density of the fine and hyperfine structure resolved rotational transition averaged over the single antenna beam.}
\tablenotetext{d}{\citet{Tenenbaum}}.
\tablenotetext{e}{ \citet{Killian}}.
\tablenotetext{f}{\citet{Caris}, and references therein.}
\tablenotetext{g}{\citet{Yamamoto}}
\tablenotetext{h}{\citet{Cerni2008}}
\tablenotetext{i}{ \citet{Gottlieb2010}}
\label{CnHspace}
\end{deluxetable}

\pagebreak

\begin{deluxetable}{ccc lrc ccl r}
\tablecaption{Measured Rotational Frequencies in the  $1v_7~(^2\Pi)$ Excited Vibrational Level of C$_4$H}
\tabletypesize{\footnotesize}
\tablewidth{0pt}
\tablehead{
 \multicolumn{1}{c}{}    
 & \multicolumn{4}{c}{$J~=~N~-~\frac{1}{2}$ }   
 & \multicolumn{1}{c}{}                      
 & \multicolumn{4}{c}{$J~=~N~+~\frac{1}{2}$ } \\
 \cline{2-5}
 \cline{7-10}
\multicolumn{1}{c}{$N'~\leftarrow~N$} & \multicolumn{1}{c}{Parity} & \multicolumn{1}{c}{$F'~\leftarrow~F$}   &
\multicolumn{1}{c}{Frequency} &  \multicolumn{1}{c}{$O-C$\tablenotemark{a}}           &
\multicolumn{1}{c}{}       &
\multicolumn{1}{c}{Parity}  & \multicolumn{1}{c}{$F'~\leftarrow~F$} & \multicolumn{1}{c}{Frequency}   &
\multicolumn{1}{c}{$O-C$\tablenotemark{a}}                             \\
 \colhead{}                                                  &  \colhead{}                                   &       \colhead{}                                                                  
& \multicolumn{1}{c}{(MHz)}                   &  \multicolumn{1}{c}{(MHz)}  
& \multicolumn{1}{c}{}                  
& \colhead{}                                              &  \colhead{}                                    & \multicolumn{1}{c}{(MHz)}  
& \multicolumn{1}{c}{(MHz)}      
}
\startdata

$ 5 \leftarrow  4$ & $f$ & $ 4 \leftarrow  3$ &  44712.367 &    0.036          &   &         &          &            &          \\  
                              & $f$ & $ 5 \leftarrow  4$ &  44712.913 &    0.007          &   &         &          &            &          \\ 
                              & $e$ & $ 4 \leftarrow  3$ &  44830.238 & $-0.018$         &   &         &          &            &          \\  
                              & $e$ & $ 5 \leftarrow  4$ &  44830.966 &    0.010            &   &         &          &            &          \\ 

$ 7 \leftarrow  6$ &     &                    &            &          &          
                   & $e$ & $ 8 \leftarrow  7$ &  68553.686 &    0.025 \\
                   &     &                    &            &          &           
                   & $e$ & $ 7 \leftarrow  6$ &  68554.209 &    0.022 \\
                   &     &                    &            &          &           
                   & $f$ & $ 8 \leftarrow  7$ &  68716.586 & $-0.008$ \\
                   &     &                    &            &          &           
                   & $f$ & $ 7 \leftarrow  6$ &  68717.020 &    0.004 \\

$ 8 \leftarrow  7$ & $f$ & $ 7 \leftarrow  6$ &  74141.426 &    0.001          & 
                   & $e$ & $ 9 \leftarrow  8$ &  77833.064 & $-0.005$ \\
                   & $f$ & $ 8 \leftarrow  7$ &  74141.841 &    0.003          & 
                   & $e$ & $ 8 \leftarrow  7$ &  77833.510 & $-0.024$ \\
                   & $e$ & $ 7 \leftarrow  6$ &  74352.312 &    0.015          & 
                   & $f$ & $ 9 \leftarrow  8$ &  78029.136 &    0.020 \\
                   & $e$ & $ 8 \leftarrow  7$ &  74352.817 & $-0.002$          & 
                   & $f$ & $ 8 \leftarrow  7$ &  78029.504 &    0.024 \\

$ 9 \leftarrow  8$ & $f$ & $ 8 \leftarrow  7$ &  83878.977 & $-0.019$          & 
                   & $e$ & $10 \leftarrow  9$ &  87142.553 &    0.005 \\
                   & $f$ & $ 9 \leftarrow  8$ &  83879.350 & $-0.003$          & 
                   & $e$ & $ 9 \leftarrow  8$ &  87142.954 & $-0.002$ \\
                   & $e$ & $ 8 \leftarrow  7$ &  84122.980 &    0.021          & 
                   & $f$ & $10 \leftarrow  9$ &  87371.839 & $-0.027$ \\
                   & $e$ & $ 9 \leftarrow  8$ &  84123.434 &    0.013          & 
                   & $f$ & $ 9 \leftarrow  8$ &  87372.159 & $-0.019$ \\
$10 \leftarrow  9$ & $f$ & $ 9 \leftarrow  8$ &  93585.949 & $-0.022$          & 
                   &     &                    &            &          \\ 
                   & $f$ & $10 \leftarrow  9$ &  93586.275 & $-0.003$          & 
                   &     &                    &            &          \\ 
                   & $e$ & $ 9 \leftarrow  8$ &  93863.168 & $-0.011$          & 
                   &     &                    &            &          \\ 
                   & $e$ & $10 \leftarrow  9$ &  93863.572 & $-0.013$          & 
                   &     &                    &            &          \\ 
\enddata
\tablenotetext{a}{Frequencies calculated with the constants in Table~\ref{tab-const}.}
\label{tab-nu7pi}
\end{deluxetable}

\pagebreak

\begin{deluxetable}{ccrr c  crr}
\tablecaption{Measured Rotational Frequencies in the  $2\nu_7~(^2\Sigma)$   Excited Vibrational Level of C$_4$H}
\tablewidth{0pt}
\tablehead{
\multicolumn{1}{c}{}  &
\multicolumn{3}{c}{$J~=~N~-~\frac{1}{2}$ } 
& \multicolumn{1}{c}{}                  
&\multicolumn{3}{c}{$J~=~N~+~\frac{1}{2}$ } \\
\cline{2-4}
\cline{6-8}
\multicolumn{1}{c}{$N'~\leftarrow~N$} &
\multicolumn{1}{c}{$F'~\leftarrow~F$} &
\multicolumn{1}{c}{Frequency} &
\multicolumn{1}{c}{$O-C$\tablenotemark{a}} &

\multicolumn{1}{c}{}                  &
\multicolumn{1}{c}{$F'~\leftarrow~F$} &
\multicolumn{1}{c}{Frequency} &
\multicolumn{1}{c}{$O-C$\tablenotemark{a}} \\

\multicolumn{2}{c}{}        
&\multicolumn{1}{c}{(MHz)} 
& \multicolumn{1}{c}{(MHz)}
&\multicolumn{2}{c}{}                   
& \multicolumn{1}{c}{(MHz)} 
& \multicolumn{1}{c}{(MHz)} 

}
\startdata

$2 \leftarrow 1$ & $2 \leftarrow 1$   & 19157.642    &    0.002            &  & $3 \leftarrow 2$   &  19100.146 &    $0.000$  \\

$3 \leftarrow 2$ & $3 \leftarrow 2$   & 28721.504   &    $-0.003$       &  & $3 \leftarrow 2$   &  28664.320 &    $-0.002$  \\
                            & $2 \leftarrow 1$   & 28721.625   &    $-0.001$       &  & $4 \leftarrow 3$   &  28664.476 &    $0.002$  \\

$4 \leftarrow 3$ & $4 \leftarrow 3$   & 38285.562   &    0.000       &  & $4 \leftarrow 3$   &  38228.637 &    $-0.002$  \\
                            & $3 \leftarrow 2$   & 38285.583   &    0.002        &  & $5 \leftarrow 4$   &  38228.734 &       0.002  \\

\enddata
\tablecomments{Estimated $1\sigma$ uncertainties in the centimeter-wave measurements is 5~kHz.}
\tablenotetext{a}{Frequencies calculated with the constants in Table~\ref{tab-const}.}
\label{tab-nu7sig}
\end{deluxetable}


\begin{deluxetable}{cclrcclr}
\tablecaption{Measured Rotational Frequencies in the  $2v_7~(^2\Delta)$   Excited Vibrational Level of C$_4$H}
\tablewidth{0pt}
\tablehead{

\multicolumn{1}{c}{}   
&\multicolumn{3}{c}{$J~=~N~-~\frac{1}{2}$ } 
& \multicolumn{1}{c}{}                   
&\multicolumn{3}{c}{$J~=~N~+~\frac{1}{2}$ } \\
\cline{2-4}
\cline{6-8}
\multicolumn{1}{c}{$N'~\leftarrow~N$} &
\multicolumn{1}{c}{$F'~\leftarrow~F$} &
\multicolumn{1}{c}{Frequency} &
\multicolumn{1}{c}{$O-C$\tablenotemark{a}} 
& \multicolumn{1}{c}{} 
&\multicolumn{1}{c}{$F'~\leftarrow~F$} &
\multicolumn{1}{c}{Frequency} &
\multicolumn{1}{c}{$O-C$\tablenotemark{a}} \\

\colhead{}   &  \colhead{}    & \multicolumn{1}{c}{(MHz)} 
& \multicolumn{1}{c}{(MHz)} 
& \multicolumn{1}{c}{} 
&  \colhead{}    & \multicolumn{1}{c}{(MHz)} 
& \multicolumn{1}{c}{(MHz)} 

}
\startdata

$ 7 \leftarrow  6$ &                    &            &            &
                   & $ 8 \leftarrow  7$ &  69192.595 &    0.020 \\
                               &                    &            &            &
                   & $ 7 \leftarrow  6$ &  69193.060 &    0.007 \\
$ 8 \leftarrow  7$ & $ 7 \leftarrow  6$ &  74161.246 &    0.017    & 
                   & $ 9 \leftarrow  8$ &  78519.431 &    0.009 \\
                   & $ 8 \leftarrow  7$ &  74161.662 &    0.019                & 
                   & $ 8 \leftarrow  7$ &  78519.898 &    0.033 \\
$ 9 \leftarrow  8$ & $ 8 \leftarrow  7$ &  83947.326 & $-$0.002 &  
                   & $10 \leftarrow  9$ &  87870.583 & $-$0.013 \\
                   & $ 9 \leftarrow  8$ &  83947.724 &    0.003                 &
                   & $ 9 \leftarrow  8$ &  87870.967 & $-$0.033 \\
$10 \leftarrow  9$ & $ 9 \leftarrow  8$ &  93708.740 &    0.001    &
                   &                    &  &    \\
                   & $10 \leftarrow  9$ &  93709.107 &    0.003                &
                   &                    &  &    \\

\enddata
\tablenotetext{a}{Frequencies calculated with the constants in Table~\ref{tab-const}.}

\label{tab-nu7delta}

\end{deluxetable}

 \pagebreak

\begin{deluxetable}{ccrr c  crr}
\tablecaption{Measured Rotational Frequencies in the $a^2\Pi$ Excited Vibrational Level of C$_4$H} 
\tabletypesize{\small}
\tablewidth{0pt}
\tablehead{

& \multicolumn{3}{c}{$J~=~N~-~\frac{1}{2}$ } 
& \multicolumn{1}{c}{}
& \multicolumn{3}{c}{$J~=~N~+~\frac{1}{2}$ } \\
\cline{2-4}
\cline{6-8}
\multicolumn{1}{c}{$N'~\leftarrow~N$} &
\multicolumn{1}{c}{Parity} &
\multicolumn{1}{c}{Frequency} &
\multicolumn{1}{c}{$O-C$\tablenotemark{a}} 
& \multicolumn{1}{c}{}
&\multicolumn{1}{c}{Parity} &
\multicolumn{1}{c}{Frequency} &
\multicolumn{1}{c}{$O-C$\tablenotemark{a}} \\

& & \multicolumn{1}{c}{(MHz)} 
& \multicolumn{1}{c}{(MHz)} 
& \multicolumn{1}{c}{}
& & \multicolumn{1}{c}{(MHz)} 
& \multicolumn{1}{c}{(MHz)} 

}
\startdata

$10 \leftarrow 9$    & $e$ &  95502.793 &    0.032 &     &$f$ &  95557.501 &    0.028\\
                                 & $f$ &  95542.080   &    0.031 &     & $e$ &  95609.099 &    0.014\\
$13 \leftarrow 12$ & $e$ & 124182.153 & $-$0.006 &  & $f$ & 124198.118 &    0.012\\
                                 & $f$ & 124235.243  & $-$0.024 &  & $e$ & 124263.526 & $-$0.008\\
$15 \leftarrow 14$ & $e$ & 143293.623 & $-$0.030 &  & $f$ & 143295.568 & $-$0.009\\
                                 & $f$ & 143356.034  & $-$0.004 &  & $e$ & 143370.299 &    0.019\\
$18 \leftarrow 17$ & $e$ & 171954.051 &    0.000 &     & $f$ & 171942.999 & $-$0.028\\
                                 & $f$ & 172030.463  & $-$0.008 &  & $e$ & 172031.797 &    0.032\\
$19 \leftarrow 18$ & $e$ & 181506.072 & $-$0.037 &  &     &            &         \\
                                 & $f$ & 181587.188  & $-$0.056 &  & $e$ & 181585.504 & $-$0.016\\
$30 \leftarrow 29$ & $e$ & 286540.183 &    0.012 &     & $f$ & 286510.241 & $-$0.079\\
                                 & $f$ & 286674.782  &    0.010 &     & $e$ & 286657.242 &    0.004\\
$38 \leftarrow 37$ & $e$ & 362881.183 &    0.024 &     & $f$ & 362847.399 &    0.015\\
                                 & $f$ & 363057.084  &    0.053 &     & $e$ & 363035.520 & $-$0.053   \\

\enddata
\tablenotetext{a}{Frequencies calculated with the constants in Table~\ref{NewConst}.}
\label{tab-pia}
\end{deluxetable}

 
\pagebreak

\begin{deluxetable}{ccc rrc crr r}
\tablecaption{Measured Rotational Frequencies in the $b^2\Pi$  Excited Vibrational Level of C$_4$H}
\tabletypesize{\footnotesize}
\tablewidth{0pt}
\tablehead{
& \multicolumn{4}{c}{$J~=~N~-~\frac{1}{2}$ } 
& \multicolumn{1}{c}{} 
& \multicolumn{4}{c}{$J~=~N~+~\frac{1}{2}$ } \\
\cline{2-5}
\cline{7-10}
\multicolumn{1}{c}{$N'~\leftarrow~N$} &
\multicolumn{1}{c}{Parity} &
\multicolumn{1}{c}{$F'~\leftarrow~F$} &
\multicolumn{1}{c}{Frequency} &
\multicolumn{1}{c}{$O-C$\tablenotemark{a}} 
& \multicolumn{1}{c}{} 
&\multicolumn{1}{c}{Parity} &
\multicolumn{1}{c}{$F'~\leftarrow~F$} &
\multicolumn{1}{c}{Frequency} &
\multicolumn{1}{c}{$O-C$\tablenotemark{a}} \\

& & & \multicolumn{1}{c}{(MHz)} 
& \multicolumn{1}{c}{(MHz)} 
& \multicolumn{1}{c}{} 
& & & \multicolumn{1}{c}{(MHz)} 
& \multicolumn{1}{c}{(MHz)} 

}
\startdata

$7 \leftarrow 6$   &     &                    &            &             &
                   & $f$ &  $8 \leftarrow 7$  &  69557.920 &    0.032 \\
                   &     &                    &            &             &
                   & $f$ &  $7 \leftarrow 6$  &  69558.338 & $-0.018$ \\
                   &     &                    &            &             &
                   & $e$ &  $8 \leftarrow 7$  &  69689.931 &    0.026 \\
                   &     &                    &            &             &
                   & $e$ &  $7 \leftarrow 6$  &  69690.356 &    0.029 \\

$8 \leftarrow 7$   & $e$ &  $7 \leftarrow 6$  &  73404.509 &    0.034             &
                   &     &                    &            &          \\
                   & $e$ &  $8 \leftarrow 7$  &  73404.924 &    0.021             &
                   &     &                    &            &          \\
                   & $f$ &  $7 \leftarrow 6$  &  73600.331 &    0.008             &
                   & $e$ &  $9 \leftarrow 8$  &  79038.104 & $-0.017$ \\
                   
                   & $f$ &  $8 \leftarrow 7$  &  73600.807 &    0.004             &
                   & $e$ &  $8 \leftarrow 7$  &  79038.541 &    0.027 \\                 
                 
$9 \leftarrow 8$   & $e$ &  $8 \leftarrow 7$  &  83138.919 &    0.008             &
                   & $f$ & $10 \leftarrow 9$  &  88203.982 &    0.003 \\
                   & $e$ &  $9 \leftarrow 8$  &  83139.341 &    0.032             &
                   & $f$ &  $9 \leftarrow 8$  &  88204.406 &    0.014 \\
                   & $f$ &  $8 \leftarrow 7$  &  83366.506 &    0.006             &
                   & $e$ & $10 \leftarrow 9$  &  88401.284 & $-0.007$ \\
                   & $f$ &  $9 \leftarrow 8$  &  83366.936 & $-0.013$             &
                   & $e$ &  $9 \leftarrow 8$  &  88401.641 & $-0.013$ \\

$10 \leftarrow 9$  & $e$ &  $9 \leftarrow 8$  &  92858.025 & $-0.003$             &
                   &     &                    &            &          \\
                   & $e$ & $10 \leftarrow 9$  &  92858.378 & $-0.019$             &
                   &     &                    &            &          \\
                   & $f$ &  $9 \leftarrow 8$  &  93118.952 & $-0.002$             &
                   &     &                    &            &          \\
                   & $f$ & $10 \leftarrow 9$  &  93119.392 &    0.019             &
                   &     &                    &            &          \\

$11 \leftarrow 10$ & $e$ & $10 \leftarrow 9$  & 102562.162 & $-0.011$             &
                   &     &                    &            &          \\
                   & $e$ & $11 \leftarrow 10$ & 102562.488 & $-0.024$             &
                   &     &                    &            &          \\
                   & $f$ & $10 \leftarrow 9$  & 102857.575 & $-0.019$             &
                   &     &                    &            &          \\
                   & $f$ & $11 \leftarrow 10$ & 102857.964 & $-0.019$             &
                   &     &                    &            &          \\

$12 \leftarrow 11$ & $e$ & $11 \leftarrow 10$ & 112251.885 & $-0.020$             &
                   & $f$ & $13 \leftarrow 12$ & 116272.310 & $-0.013$ \\
                   & $e$ & $12 \leftarrow 11$ & 112252.194 & $-0.022$             &
                   & $f$ & $12 \leftarrow 11$ & 116272.661 &    0.009 \\
                   & $f$ & $11 \leftarrow 10$ & 112582.623 &    0.022             &
                   &     &                    &            &          \\
                   & $f$ & $12 \leftarrow 11$ & 112582.992 &    0.020             &
                   &     &                    &            &          \\

$13 \leftarrow 12$  &       &                 &            &             &           
                                 & $f$ &                  & 125653.515 & $-0.011$ \\
                                 &       &                  &            &             &           
                                 & $e$ &                & 125992.970 & $-0.028$ \\
 
$15 \leftarrow 14$ & $e$ &        & 141242.289 & $-0.005$             &  
                   & $f$ &                    & 144448.893 & $-0.030$ \\
                   & $f$ &                    & 141680.954 &    0.009            &  
                   & $e$ &                    & 144860.942 &    0.014 \\

$16 \leftarrow 15$ &     &                    &            &              &         
                   & $f$ &                    & 153861.682 & $-0.015$ \\   
                   & $f$ &                    & 151356.927 &    0.00              & 
                   & $e$ &                    & 154309.767 & $-0.036$ \\

$18 \leftarrow 17$ & $e$ &         & 170131.947 & $-0.045$             &  
                   & $f$ &                    & 172713.383 & $-0.026$ \\
                   & $f$ &                    & 170678.204 & $-0.054$             & 
                   & $e$ &                    & 173232.979 & $-0.058$    \\
$19 \leftarrow 18$ & $e$ &        & 179743.519 &    0.073             &  
                   & $f$ &                    & 182150.890 &    0.071 \\
                   & $f$ &                    & 180325.149 &    0.033              & 
                   & $e$ &                    & 182705.875 &    0.078 \\

$25 \leftarrow 24$ & $e$ &        & 237266.740 & $-0.031$              & 
                   & $f$ &                    & 238895.750 & $-0.025$ \\
                   &     &                      &            &              &         
                   & $e$ &                   & 239656.612 &    0.018 \\

$30 \leftarrow 29$ & $e$ &        & 285066.869 & $-0.047$             &  
                   & $f$ &                    & 286286.672 & $-0.019$ \\
                   & $f$ &                    & 286016.697\tablenotemark{b}  &             &
                   & $e$ &                    & 287211.598 & $-0.015$ \\

$38 \leftarrow 37$ & $e$ &        & 361388.690 &    0.044              & 
                   & $f$ &                    & 362204.016 & $-0.001$ \\
                   & $f$ &                    & 362584.458\tablenotemark{b} &             &
                   & $e$ &                    & 363380.662 &    0.000 \\

\enddata
\label{tab-pib}
\tablenotetext{a}{Frequencies calculated with the constants in Table~\ref{NewConst}.}
\tablenotetext{b}{Not included in the fit owing to the perturbation (see Section~\ref{subsec:bSigma}).}
\end{deluxetable}
      
\pagebreak

\begin{deluxetable}{ccrr c  crr}
\tablecaption{Measured Rotational Frequencies in the $a^2\Sigma $ Excited Vibrational Level of C$_4$H}
\tablewidth{0pt}
\tablehead{
\multicolumn{1}{c}{}  &
\multicolumn{3}{c}{$J~=~N~-~\frac{1}{2}$ } 
& \multicolumn{1}{c}{}                  
&\multicolumn{3}{c}{$J~=~N~+~\frac{1}{2}$ } \\
\cline{2-4}
\cline{6-8}
\multicolumn{1}{c}{$N'~\leftarrow~N$} &
\multicolumn{1}{c}{$F'~\leftarrow~F$} &
\multicolumn{1}{c}{Frequency} &
\multicolumn{1}{c}{$O-C$\tablenotemark{a}} &

\multicolumn{1}{c}{}                  &
\multicolumn{1}{c}{$F'~\leftarrow~F$} &
\multicolumn{1}{c}{Frequency} &
\multicolumn{1}{c}{$O-C$\tablenotemark{a}} \\

\multicolumn{2}{c}{}        
&\multicolumn{1}{c}{(MHz)} 
& \multicolumn{1}{c}{(MHz)}
&\multicolumn{2}{c}{}                   
& \multicolumn{1}{c}{(MHz)} 
& \multicolumn{1}{c}{(MHz)} 

}
\startdata

$2 \leftarrow 1$ & $2 \leftarrow 1$   & 19052.195    &    0.002            &  & $3 \leftarrow 2$   &  19034.551 &    $0.000$  \\
                            &                             &                      &                         &  & $2 \leftarrow 1$   &  19034.559 &    $0.000$  \\

$3 \leftarrow 2$ & $3 \leftarrow 2$   & 28573.715   &    $-0.003$       &  & $4 \leftarrow 3$   &  28556.273 &    $-0.002$  \\
                            & $2 \leftarrow 1$   & 28574.092   &    $-0.001$       &  & $3 \leftarrow 2$   &  28556.305 &    $0.002$  \\

$4 \leftarrow 3$ & $4 \leftarrow 3$   & 38095.249   &    $0.000$       &  & $5 \leftarrow 4$   &  38077.887 &    $-0.002$  \\
                            & $3 \leftarrow 2$   & 38095.401   &       0.002        &  & $4 \leftarrow 3$   &  38077.915 &       0.002  \\
$15 \leftarrow 14$ &                         & 142822.059 &    0.059            &  &                             &  142805.027 &    0.017  \\
$16 \leftarrow 15$ &                         & 152341.151 &    0.036          &  &                             &  152324.186 &    0.023  \\
$18 \leftarrow 17$ &                         & 171378.323 &    0.027          &  &                             &  171361.426 & $-0.001$  \\ 
$25 \leftarrow 24$ &                         &  237995.258 &   0.061           &  &                             &  237978.662 & $-0.036$  \\
$30 \leftarrow 29$ &                         &  285563.150 &   $-0.007$           &  &                             &  285546.928 & $-0.068$  \\
$38 \leftarrow 37$ &                         &  361637.717 & $-0.039$         &  &                             &  361622.319 &    0.057  \\

\enddata
\tablecomments{Estimated $1\sigma$ uncertainties in the centimeter-wave measurements is 5~kHz.}
\tablenotetext{a}{Frequencies calculated with the constants in Table~\ref{NewConst}.}
\label{tab-sia}
\end{deluxetable}


\pagebreak

\begin{deluxetable}{ccrr c  crr}
\tablecaption{Measured Rotational Frequencies in the $b^2\Sigma $ Excited Vibrational Level of C$_4$H}   
\tablewidth{0pt}
\tablehead{
\multicolumn{1}{c}{}  &
\multicolumn{3}{c}{$J~=~N~-~\frac{1}{2}$ } 
& \multicolumn{1}{c}{}                  
&\multicolumn{3}{c}{$J~=~N~+~\frac{1}{2}$ } \\
\cline{2-4}
\cline{6-8}
\multicolumn{1}{c}{$N'~\leftarrow~N$} &
\multicolumn{1}{c}{$F'~\leftarrow~F$} &
\multicolumn{1}{c}{Frequency} &
\multicolumn{1}{c}{$O-C$\tablenotemark{a}} &

\multicolumn{1}{c}{}                  &
\multicolumn{1}{c}{$F'~\leftarrow~F$} &
\multicolumn{1}{c}{Frequency} &
\multicolumn{1}{c}{$O-C$\tablenotemark{a}} \\

\multicolumn{2}{c}{}        
&\multicolumn{1}{c}{(MHz)} 
& \multicolumn{1}{c}{(MHz)}
&\multicolumn{2}{c}{}                   
& \multicolumn{1}{c}{(MHz)} 
& \multicolumn{1}{c}{(MHz)} 

}
\startdata

$2 \leftarrow 1$ &                              &                  &                      &  & $3 \leftarrow 2$   &  19000.323 &    0.005  \\

$3 \leftarrow 2$ & $2 \leftarrow 1$   & 28558.709&    $-0.002$     &  & $3\leftarrow 2$   &  28511.978 &    $-0.002$  \\
                            &   &  &             &  & $4 \leftarrow 3$   &  28512.176 &       0.007  \\

$4 \leftarrow 3$ & $3 \leftarrow 2$   & 38070.179 &    $0.012$     &  & $4 \leftarrow 3$   &  38023.823 &    $-0.004$  \\
                            & $4 \leftarrow 3$   & 38070.204 &   $-0.008$     &  & $5 \leftarrow 4$   &  38023.924 &     $-0.004$  \\

$ 8 \leftarrow  7$   &    &  76114.910   &    0.035     &       &    &                      &    \\
$ 9 \leftarrow  8$   &    &  85625.489   &    $-0.018$     &       &    &  85580.757 &    $-0.037$ \\
$10 \leftarrow  9$  &    &  95135.697   & $-0.108$  &       &    &  95091.505 & $-0.118$   \\
$11 \leftarrow 10$ &    & 104645.517  & $-0.168$  &       &    & 104602.169 &  $-0.049$   \\
$12 \leftarrow 11$ &    & 114154.716  & $-0.302$ &       &    & 114112.540  &   $-0.016$   \\
$13 \leftarrow 12$ &    &                        &                  &       &    & 123622.556 & $-0.058$   \\
$15 \leftarrow 14$ &    &                        &                  &       &    & 142641.738 & $-0.057$  \\
$16 \leftarrow 15$ &    &                        &                  &       &    & 152150.770 & $-0.103$   \\
$18 \leftarrow 17$ &    &                        &                  &       &    & 171167.819 & $-0.067$   \\
$19 \leftarrow 18$ &    &                        &                  &       &    & 180675.774 & $-0.002$   \\
$24 \leftarrow 23$ &    &                        &                  &       &    & 228208.117 & $-0.022$   \\
$25 \leftarrow 24$ &    & 237750.637 & $-0.119$ &        &    & 237713.112 &    0.077   \\
$27 \leftarrow 26$ &    & 256763.157 &    0.093    &        &    & 256720.966 & $-0.104$   \\
$30 \leftarrow 29$ &    & 285272.924 &    0.168   &        &    & 285228.362 & $-0.030$   \\
$34 \leftarrow 33$ &    & 323274.033 &    0.131    &        &    & 323228.432 &    0.044   \\
$35 \leftarrow 34$ &    &                       &                  &        &    & 332726.510 &    0.011   \\
$36 \leftarrow 35$ &    & 342269.653 & 0.050 &        &    & 342223.785 &    $-0.024$   \\
$38 \leftarrow 37$ &    & 361261.840 & $-0.090$ &        &    & 361215.873 &   $-0.069$   \\

\enddata
\tablecomments{Estimated $1\sigma$ uncertainties in the centimeter-wave measurements is 5~kHz.}
\tablenotetext{a}{Frequencies calculated with the constants in Table~\ref{NewConst}.}
\label{tab-sib}

\end{deluxetable}


\pagebreak

\begin{deluxetable}{lcc cc}
\tablecaption{Spectroscopic Constants of Vibrationally Excited C$_4$H (Prior work)}
\tablewidth{0pt}
\tablehead{

\multicolumn{1}{c}{Constant}
& \multicolumn{1}{c}{ground $^2\Sigma$\tablenotemark{a}}
& \multicolumn{1}{c}{$1\nu_7~(^2\Pi)$\tablenotemark{b}}
& \multicolumn{1}{c}{$2\nu_7~(^2\Sigma)$\tablenotemark{c}}
& \multicolumn{1}{c}{$2\nu_7~(^2\Delta)$\tablenotemark{d}}                 \\

   
}
\startdata

$A$ &                & $-$89812.6(10) &                 & $-$50575.69(39)\\ 
$B$ & 4758.6557(7)   & 4762.85047(26)  & 4782.16713(28)   & 4778.57763(32)\\ 
$D \times 10^3$ & 0.8627(10) & 0.89320(41) & 0.91475(57)  & 0.90125(21)\\
$H \times 10^9$ &    & 0.93(24)        & $-2.82(38)$     & \\ 
$\gamma$ & $-$38.555(2) & $-$38.149(38) & $-56.6483(68)$  & $-$37.225(17)\\
$\gamma_D \times 10^3$ & 0.127(9)  & 0.030(14)  & 0.709(26)       & \\
 \multicolumn{5}{l}{$K$-type doubling constants:}  \\
$p$ &                & $-18.243(15)$      & & \\ 
$p_D \times 10^3$ &  & 0.1809(95)   & & \\ 
$q$ &                & $-14.96761(52)$    & & \\ 
$q_D \times 10^3$ &  & 0.11794(36)  & & \\ 
 \multicolumn{5}{l}{Hyperfine constants:}  \\
$a$                                       &                           & 1.62(49)                &                              & 1.25(34) \\ 
$b_F$\tablenotemark{e} &$-14.493$      & $-10.3(15)$        & $-14.62(20)$      & $-20.3(28)$      \\ 
$c$                                      & 12.435(10)     & 13.6(16)                &   12.29(21)         & 11.1(1.9) \\ 
$d$                                     &                          & 5.07(72)                &                              & \\ 
 \multicolumn{5}{l}{Standard deviation:}   \\ 
                                           &                        & 0.020                     & 0.023                   & 0.020 \\ 

\enddata
\label{tab-const}
\pagebreak
\tablecomments{Units are MHz and uncertainties (in parentheses) are 1$\sigma$.  The correlation coefficient matrices are available from the authors.}
\tablenotetext{a}{From \citet{Gottlieb1983}.}
\tablenotetext{b}{Constants derived from the measurements in \citet{Yamamoto} and Table~\ref{tab-nu7pi}.}
\tablenotetext{c}{Constants derived from the measurements in \citet{Guelin1987}, \citet{Yamamoto}, and Table~\ref{tab-nu7sig}.}  
\tablenotetext{d}{Constants derived from the measurements in \citet{Yamamoto} and Table~\ref{tab-nu7delta}.}
\tablenotetext{e}{ $b_F~=~b~+~\frac{1}{3}c$.}
\end{deluxetable}

\pagebreak

\begin{deluxetable}{lcc cc}
\tablecaption{Spectroscopic Constants of Vibrationally Excited C$_4$H (This work)}
\tablewidth{0pt}
\tablehead{

\multicolumn{1}{c}{Constant\tablenotemark{a}}
& \multicolumn{1}{c}{$a^2\Pi$ }
& \multicolumn{1}{c}{$b^2\Pi$ }
& \multicolumn{1}{c}{$a^2\Sigma$}
& \multicolumn{1}{c}{$b^2\Sigma$}   \\


}
\startdata

$A$                                          & $-7011.1(25)$ & $-$152601.3(69) &    &      \\
$B$                                          & 4778.03697(39) & 4770.99746(38) & 4760.85127(32) & 4755.9431(14) \\
$D \times 10^3$                    & 0.80158(87)    & 0.91582(76)    & 0.89191(62)     & 0.9762(17) \\
$H \times 10^9$                    & 1.04(36)       & 11.65(33)      & 2.28(28)       & 3.94(72) \\
$\gamma$                              & $-$34.371(32)  & $-$28.94(19)   & $-$17.266(8)  & $-$46.674(33) \\
$\gamma_D \times 10^3$   &         & 1.653(62)		  & 0.409(10)    & 0.038(18) \\
 \multicolumn{5}{l}{$K$-type doubling constants:}  \\
$p$                                          & $-12.317(21)$  & $-31.716(28)$     & & \\
$p_D \times 10^3$               &              & $-1.433(48)$   & & \\
$q$                                         &2.26473(50)    & $-16.2750(11)$    & & \\
$q_D \times 10^3$              & 0.04516(36)  & 0.1889(20)     & & \\
$q_H \times 10^9$              &   & $-10.42(87)$    & & \\
 \multicolumn{5}{l}{Coriolis constants\tablenotemark{a}:}   \\
$q_l$                                     &                 &                                        &                                        & $-0.0728(15)$ \\
$\rho$ $\times 10^3$          &                  &                                        &                                        & $-2.5981(99)$ \\
 \multicolumn{5}{l}{Hyperfine constants:}   \\
$a$                                        &                & 0.0\tablenotemark{b}   &                                           &                                                 \\ 
$b_F$\tablenotemark{c}     &                & $-7.4(18)$                     &   $-5.07(13)$                    &   $-9.27(83)$                           \\
$c$                                        &                & 18.9(22)                        &     13.65(25)                      &      6.0(8)                                    \\
$d$                                        &                & 3.2(20)                           &                                             &                                                \\
 \multicolumn{5}{l}{Standard deviation:}   \\ 
                                            & 0.032        & 0.034                             & 0.030                         & 0.035 \\
\enddata
\label{NewConst}
\tablecomments{Units are MHz and uncertainties (in parentheses) are 1$\sigma$.  The correlation coefficient matrices are available from the authors.}
\tablenotetext{a}{For a discussion of the constants see Section~\ref{sec:perturbations}, and for the sign conventions see Section~\ref{sec:LabSpectroscopic}.}
\tablenotetext{b}{Constrained to zero.}
\end{deluxetable}


\pagebreak

\begin{deluxetable}{lr ccr }
\tablecaption{C$_4$H Vibrational Frequencies (in cm$^{-1}$).}
\tablewidth{0pt}
\tablehead{

\multicolumn{1}{l}{Mode} 
& \multicolumn{1}{c}{}
& \multicolumn{1}{c}{New levels}
& \multicolumn{1}{c}{Laser\tablenotemark{a}}
& \multicolumn{1}{c}{Theoretical\tablenotemark{b} }   \\

\multicolumn{2}{c}{}
& \multicolumn{1}{c}{(This work)} 
&  \multicolumn{2}{c}{}


 }
 
\startdata


$\nu_7$                &     &                             &   168             &       178                      \\          
$\nu_6$                &     &   $b ^2\Pi$         &                       &       379                       \\          
$\nu_5$                &     &   $a ^2\Pi$         &                       &       578                      \\           
$\nu_4$                &     &                             &    930            &       924                      \\           
$\nu_3$                &    &                              &                       &    2131                      \\           
$\nu_2$                &    &                              &                       &    2301                       \\          
$\nu_1$                &    &                              &                       &    3608                       \\          
$\nu_6 + \nu_7$ &    &  $a^2\Sigma$    &     565           &     557                         \\         
$\nu_5 + \nu_7$ &    &  $b^2\Sigma$    &                       &     756                        \\          

\enddata

\tablenotetext{a}{Frequenices derived from the electronic spectrum of C$_4$H using double resonance four-wave mixing \citep{Mazzotti}.}
\tablenotetext{b}{Harmonic vibrational frequencies at the MCSCF/cc-pVQZ level of theory \citep{Graf}.}

\label{tab-hc3n}
\end{deluxetable}



\end{document}